\documentstyle[aps,preprint,tighten,epsfig]{revtex}

 \input myepic.sty
 \input myeepic.sty

\def\xara(#1,#2,#3,#4){\left(\matrix{#1 &
 #2\cr #3 & #4\cr}\right)}
\def\lll(#1,#2){\left(\matrix{#1 \cr #2 \cr}\right)}
\def\rrr(#1,#2){\left(\matrix{#1 & #2 \cr}\right)}

\newcommand{\del}{{\bf\nabla}}
\def\It{\tau}
 \def\equ(#1){Eq.\ (\ref{#1})}
 \def\T{{\cal T}}
 \def\Re{\mbox{Re}}
 \def\Im{\mbox{Im}}
\def\Arg{\mbox{Arg}}
\def\nnn{\nonumber}

\def\G5{K}

\def\finv{F^{-1}}
\def\rt{\Re T }
\def\rk{\Re K }
\def\imt{\Im T }
\def\ik{\Im K }
\def\I{{\cal I}}
\def\J{{\cal J}}
\def\Ip{ {\cal I}_{\pi} }
\def\Ie{ {\cal I}_{\eta} }
\def\dpl{\delta_{\pi}^l}
\def\del{\delta_{\eta}^l}
\def\tpp{ {1\over 2i} (\rho_l e^{2 i \dpl}-1) }
\def\tee{ {1\over 2i} (\rho_l e^{2 i \del}-1) }
\def\tpe{ {1\over 2} \sqrt{1-\rho_l^2} e^{i(\dpl+\del)}  } 

\def\lpp{\lambda_{\pi\pi}}
\def\lee{\lambda_{\eta\eta}}
\def\lpe{\lambda_{\pi\eta}}
\def\lep{\lambda_{\eta\pi}}
\def\thru#1{\mathrel{\mathop{#1\!\!\!/}}}
\def\mmm{ {\cal{M}}}
\def\bea{\begin{eqnarray}}
\def\eea{\end{eqnarray}}
\def\be{\begin{equation}}
\def\ttt{\theta(\Delta_i)}
\def\ee{\end{equation}}

\def\gp{g_{\pi}}
\def\get{g_{\eta}}

\def\f{\phi}
\def\F{\Phi}
\def\dm{\partial_{\mu}}

\def\dM{\partial^{\mu}}

\def\g5{\gamma_5}
\def\L{\Lambda}
\def\tp{\thru{\partial}}

\def\pmb#1{\setbox0=\hbox{$#1$}%
\kern-.025em\copy0\kern-\wd0
\kern.05em\copy0\kern-\wd0
\kern-.025em\raise.0433em\box0 }

\begin{document}  

\title {\bf Sidewise dispersion relations and the structure 
            of the nucleon vertex}

\author{R.M. Davidson and  G.I. Poulis}

\address{National Institute for Nuclear Physics and
 High-Energy 
Physics (NIKHEF)\\
P.O. Box 41882, NL-1009 DB Amsterdam, The Netherlands \\[2ex]}
\medskip
%%% \date{\today}
%%% \preprint{NIKHEF 95-058}

\maketitle

\begin{abstract}{\em }
 
We revisit sidewise dispersion relations as a method to relate
the nucleon off-shell form factor to observable quantities, namely
the meson-nucleon scattering phase shifts. It is shown how for
meson-nucleon scattering a redefinition of the intermediate fields leaves 
the scattering amplitude invariant,
but changes the behavior of the off-shell form factor as
expressed through dispersion relations, thus showing representation
dependence. We also employ a coupled-channel 
unitary model to test the validity of approximations concerning the 
influence of inelastic channels in the sidewise dispersion relation method.

\end{abstract}
\draft

\section{introduction}

In the theoretical description of processes at intermediate energies,
the structure of hadrons is often described by multiplying
the point-like vertex operators by form factors. It is common practice to 
assume that these vertices, {\it i.e.} their operator structures and the 
associated form factors, are in all situations the same as  
for a free on-shell hadron. This is done, for example, in the description of 
electron-nucleus scattering or in two-step reactions on a free nucleon, such 
as Compton scattering, where one is dealing with an intermediate nucleon not 
on its mass shell. In these cases, however, the 
electromagnetic vertices can have a 
much richer structure: there can be more independent vertex operators
and the form factors can depend on more than one scalar variable. The common 
treatment of such ``off-shell'' effects is to presume them small and to ignore 
them by using the free vertices. However, as much of the present effort in 
intermediate energy physics focusses on delicate effects, such as evidence of 
quark/gluon degrees of freedom or small components in the hadronic wavefunction,
it is mandatory to examine these issues in detail.

One theoretical tool for the description of the off-shell vertex of the nucleon
is the method of sidewise dispersion relations. The ``sidewise'' here 
indicates that one uses { the} method to get at the dependence
of the form factors on the invariant mass of the nucleon rather 
than, {\it e.g.} the $t$-channel four-momentum transfer. It 
has been used, {\it e.g.} for the electromagnetic form factors of the 
nucleon~\cite{Bincer,hare1,Bos}, electromagnetic transition form 
factors~\cite{bluv}, the nucleon axial-vector coupling constant~\cite{suura}, 
and the pion-nucleon form factor~\cite{Epstein}.
If one wants to calculate the half-off-shell 
$\pi NN$ form factor, knowledge of 
its phase along the cut in the energy plane is sufficient to determine it via 
a sidewise dispersion integral. Below the two pion 
threshold this phase is given in terms of the known pion-nucleon phase shifts. 
However, above this threshold assumptions have been made for the 
phase~\cite{Bincer,Treiman} which lead to quite different predictions for the 
half-off-shell form factor. Since neither of these prescriptions has been
tested, we use a coupled-channel, unitary model to investigate the validity of 
the assumptions regarding the phase of the off-shell meson-nucleon form factor.

Another objective of this work is to investigate  
the ``representation dependence'' of off-shell effects, and specifically how 
this representation dependence enters the sidewise dispersion 
relations analysis.
Off-shell vertices are described within the framework of the reduction 
formalism~\cite{LSZ} using interpolating (interacting) fields for the off-shell 
nucleon. The choice of this interpolating field is not unique. It is well 
known that on-shell S-matrix elements are oblivious to the choice
of interpolating field: different unitarily-equivalent Lagrangians 
constitute different representations of the theory and physically 
measurable quantities such as on-shell amplitudes are representation-independent
in accord with the Coleman-Wess-Zumino theorem \cite{Coleman}. In fact, 
the transformation need not be unitary; any reversible field redefinition will 
leave the on-shell amplitudes unchanged ~\cite{Kamefuchi}.
However, different interpolating fields in general lead to different off-shell
extrapolations~\cite{Coleman_book} and therefore off-shell form factors cannot 
be uniquely determined. This was recently 
demonstrated~\cite{Rudy,Stefan_Compton} 
in the framework of chiral perturbation theory. It was shown how the 
off-shell electromagnetic form factor of the pion changes under a unitary 
transformation of the Lagrangian which leaves, {\it e.g.} the Compton amplitude 
unchanged. While the {\it total} amplitude for the on-shell pion is 
representation independent, and certainly observable, the individual 
contributions from ``pole'' and contact terms are not. In other words, 
representation-dependent  ``off-shell effects'' in pole contributions 
in one representation appear as contact terms~\cite{Stefan_Compton} in another 
representation.

One now faces the following puzzle: although the off-shell form factors
are not unique and not measurable, it {\it appears} that through sidewise 
dispersion relations they can be determined from physical quantities such as 
meson-nucleon phase shifts. However, we will show below how representation 
dependence appears in the sidewise dispersion relations, making a unique 
determination of the half-off-shell $\pi NN$ form factor impossible.
This is in contrast to the use of dispersion relations for the 
determination of the pion-nucleon scattering amplitude at the 
non-physical point $\nu$ = $t$ = 0 \cite{hoe}, a quantity crucial
in determining the pion-nucleon sigma term \cite{gas}.

The outline of this paper is as follows: in section II we discuss
the general features of an off-shell vertex and of the representation 
dependence. In section III we review the sidewise dispersion relations and
the different assumptions proposed in the literature about their use
at energies above an inelastic threshold. These assumptions are tested in 
section IV in a simple unitary, coupled-channel model. A summary and our 
conclusions are given in section V. { Some  details of our 
calculations are contained in appendices.}

\section{The Vertex of an Off-Shell Nucleon and field redefinitions}

The most general pion-nucleon vertex, where 
the incoming nucleon of mass $m$ has momentum $p_\mu$,
the outgoing nucleon has momentum $p'_\mu$ and the pion has 
momentum $q_\mu=p'_\mu-p_\mu$, can be written as~\cite{Kazes}
\bea\label{prwth}
    \Gamma^5(p',p) &=& \Bigl[\g5 G_1 + \g5{\thru p-m\over m}G_2 
    + {\thru p'-m\over m}\g5 G_3 \nnn\\              
    && \qquad\qquad + {\thru p'-m\over m}\g5{\thru p-m\over m}
     G_4\Bigr] \ .
\eea
By sandwiching $\Gamma^5$ between on-shell spinors
     one obtains $G_1(q^2,m^2,m^2)\bar{u}(p')\gamma_{5}u(p)$.
Clearly, the off-shell vertex has much richer
     structure, in that there are 
     more independent operators and, moreover, each of them depends on more 
    kinematical variables than just the four momentum transfer, $q^2$.

Below, we will for simplicity only consider the ``half-off-shell'' 
vertex, { with the incoming nucleon on-shell}. 
Defining $w'=\sqrt{p'^2}$ and introducing
the projection operators
\be\label{prop}
P_{\pm} = { w^{\prime} \pm \thru{p}^{\prime} \over
2 w^{\prime} } \; ,
\ee
we obtain in that case
\be\label{split}
\Gamma^5(p',p) u(p) = \Bigl[ P_{+} K(q^2, w') +  
P_{-} K(q^2, -w') \Bigr] \gamma_5 u(p)  \; .
\ee
Due to the the incoming on-shell nucleon spinor, the terms proportional
to $G_2$ and $G_4$ do not contribute. The function K is obtained as
\be\label{kg}
K(q^2,\pm w')=G_1(q^2,w'^2,m^2) + {\pm w'-m\over m}
G_3(q^2,w'^2,m^2) \ .
\ee

The general electromagnetic vertex of the nucleon is more 
complicated~\cite{Bincer}. Its general form is
\be\label{elm}
\Gamma^\mu=\sum_{j,k=0,1} (\thru p')^j\Bigl[ A_1^{jk}\gamma^\mu
+ A_2^{jk} \sigma^{\mu\nu} q_\nu + A_3^{jk}q^\mu\Bigr] (\thru p)^k \ ,
\ee
where the 12 form factors, $A^{jk}_i$, are again functions of three scalar 
variables, usually taken to be $q^2$, $p^2$, and $p'^2$. By using the 
constraints 
provided by the Ward-Takahashi identity, the number of independent form factors 
can be shown to reduce to 8. Upon evaluating the vertex between two on-shell 
spinors, one recovers the familiar form of the electromagnetic current of a 
free nucleon, involving two independent contributions
with their associated form factors, such as the Dirac and Pauli form factors.
It is important to stress that in calculations of electromagnetic
reactions involving bound nucleons or two step reactions on a free nucleon,
such as Compton scattering or meson electroproduction, one
necessarily deals with the electromagnetic current or vertex of
an off-shell nucleon.

In this situation, it has been quite common to make use of ad-hoc
assumptions which use as much as possible the on-shell information
while maintaining current conservation.
Most widely used is the prescription introduced by de Forest~\cite{deForest}
for the off-shell electromagnetic current. It allows one to use the free 
current by changing its kinematical variables according to the off-shell 
situation. Another often used version for the nucleon vertex operator
was introduced by
Gross and Riska \cite{gross} and is given by
\be\label{crazy}
\Gamma_{\mu} (q^2) = \gamma_{\mu}F_1(q^2)
+ { q_{\mu} \thru{q} \over q^2} \left[ 1-F_1(q^2) \right] + \sigma_{\mu\nu} 
q^{\nu} {F_2(q^2) \over 2 m} \; .
\ee
It also only involves on-shell information, the free Dirac and Pauli
form factors, $F_1$ 
and $F_2$, but has a more general Dirac structure. The second term on the 
right-hand side of Eq. (\ref{crazy}) vanishes when the vertex is evaluated 
between on-shell spinors, but contributes when one or both nucleons
are off-shell. It
is easily seen that this vertex satisfies the Ward-Takahashi identity when free
Feynman propagators are used for the nucleon. In pion electroproduction, this 
prescription is equivalent to adding a contact term
to the Born amplitude which is 
needed to restore gauge invariance \cite{koch}.
The validity of this and other recipes can only be assessed on the 
basis of a realistic microscopic calculation and will depend on the 
kinematics of the process.

Several attempts have been made to calculate 
properties of the off-shell vertices and to estimate their effects 
in processes at intermediate energies. Bincer~\cite{Bincer}, 
for example, proposed using sidewise
dispersion relations in which the electromagnetic and strong nucleon off-shell 
form factors were related to pion-nucleon phase shifts (see section III). This 
approach was used by Nyman \cite{nyman} and Minkowski and Fisher \cite{fish}. 
Studies in the context of meson loop models have been performed, {\it e.g.} 
in Refs.~\cite{Naus,TT,mac,bos2}. Typically
effects of the order of  $5-15\%$  were found for the dependence of the form 
factors on the variable $w'$.

Recently, the off-shell pion electromagnetic vertex was investigated in the 
framework of meson chiral perturbation theory~\cite{Rudy,Stefan_Compton}. 
The computation was performed by using two different chiral Lagrangians, 
related through a unitary transformation of the fields, which leaves the 
observables unchanged. It was shown explicitly how in the description of Compton scattering
off a pion the off-shell form factors are not the same while the observable on 
shell form factor and the amplitude are the same in the two representations. 
This general result concerning the representation dependence of the off-shell
effects can be illustrated by the following simple example for pion-nucleon 
scattering. We consider the pseudoscalar meson-nucleon Lagrangian
\be\label{tria1}
    L_{ps} = {1\over 2}( %%% \dm\f\dM\f 
(\partial\phi)^2-\mu^2 \f^2) + \bar\psi(i
    \thru{\partial}- m)\psi - ig\bar\psi\g5\f\psi \ ,
\ee
and perform the transformation 
\be\label{Dyson}
   \psi \rightarrow \exp ( i \beta\g5\f) \psi \ .
\ee
Then, up to and including order $\beta^2$ terms, the new Lagrangian reads
\bea\label{triaaa3}
    \tilde{L} =  L_{ps} &-& 2 i m\beta\bar\psi\g5\f\psi 
    + \beta \bar\psi\g5(\thru\partial\f)\psi \nnn\\
    &+& 2\beta(g+2m\beta)\bar\psi\f^2 \psi
    +{\cal O}(\beta^3) \ .
\eea
This transformed Lagrangian has both pseudoscalar (PS) and pseudovector (PV) 
$\pi NN$ interaction terms, as well as a contact term. Choosing $\beta=-g/2m$
corresponds to the ``Dyson''  transformation~\cite{Friar} and the resulting
$\pi NN$ coupling becomes purely PV. 
For our discussion of the representation dependence we leave $\beta$ free
and show that physical, observable quantities are
$\beta$-independent~\cite{Harolds_suggestion}.

The meson-nucleon vertex at the tree level for both representations is readily
obtained (the dressed vertex at the one-loop level will be discussed in 
Appendix B). From $L_{ps}$ we find for the vertex
\be\label{tria3}
   \Gamma^5(p',p)_{ps} = g \g5 \ , 
\ee
corresponding to the trivial half-off-shell vertex function
\be\label{kw1}
   K(q^2, \pm w') = g \ .
\ee
On the other hand, $\tilde{L}$ yields
\bea
   \tilde{\Gamma}^5(p',p) &=&  \g5 \Bigl[g+2m\beta + \beta
   (\thru p' - \thru p ) \Bigr] 
    \label{tria4} \ ,
\eea
{ corresponding to the half-off-shell vertex function (cf. \equ(kg))}
\be\label{kw2}
{\tilde K}(q^2, \pm w') = g + \beta(m \mp w') \ ,
\ee
which is $\beta$-dependent and clearly has a different off-shell 
behavior. However, the on-shell matrix element of the vertex 
operator is the same for both representations.

What happens if we consider a two-step process on a free nucleon, such 
as pion-nucleon scattering, that involves the propagation
of an intermediate off-shell nucleon? Since this is an overall
on-shell process, the total amplitude must be independent of the value one
chooses for $\beta$. This means that the $\beta$-dependent contributions from 
the off-shell vertices in the pole terms, {\it i.e.} in the contributions 
involving two $\pi NN$ vertices connected by an intermediate nucleon 
propagator, must be compensated by some other $\beta$-dependent
contribution. To show that, we consider the {\it on-shell} 
pion-nucleon scattering T-matrix
at the tree level. Using $L_{ps}$, it involves pole terms only and reads 
\be\label{n1}
{\cal T}_{ps} = -i g^2 \bar u(p')\biggl\{ {1\over \thru p +\thru q - m}+
{1\over \thru p -\thru q' - m} \biggr\} u(p) \ ,
\ee
where $p$ and $p'$ are the intial and final nucleon four-momenta,
and $q$ and $q'$ are the intial and final pion four-momenta, respectively.
The pole term contribution to the T-matrix for the  mixed PS and PV Lagrangian,
$\tilde{L}$, is at tree level
\bea\label{nea}
    \tilde{{\cal T}}_{pole} &=& i \bar u(p')\biggl\{ 
    [g+\beta(2m + \thru q')]\g5 
    {1\over \thru p + \thru q - m} 
    \g5 [g+\beta (2m + \thru q)] \nnn \\
   +[g\!&\!+\!&\! \beta(2m -\thru q)]\g5   
     {1\over \thru p - \thru q' - m} 
    \g5 [g+\beta (2m -  \thru q')]\biggr\} u(p) \ ,
\eea
where the terms in the square brackets arise from the transformed vertex, 
Eq. (\ref{tria4}). Using the Dirac equation for the on-shell spinors, this may 
be cast in the form
\be\label{n2}
\tilde{{\cal T}}_{pole} = -i g^2 \bar u(p')
\biggl\{ {1\over \thru p +\thru q - m}+
{1\over \thru p -\thru q' - m}+4\beta(g+m\beta)\biggr\} u(p) \ ,
\ee
where the $\beta$-dependent term reflects the different ``off-shell''
behavior of the vertex obtained from $\tilde{L}$.
However, there is now also a contribution from the contact term in $\tilde{L}$,
the term proportional to $\bar\psi\f^2 \psi$, which yields 
\be\label{cont}
\tilde{{\cal T}}_{contact} = i\bar u(p')\Bigl\{ 
4\beta(g+m\beta)\Bigr\} u(p) \ .
\ee
Clearly, the $\beta$-dependent terms cancel and the total amplitude remains 
unchanged,
\be
{\cal T}_{ps} = \tilde{{\cal T}}_{pole}+\tilde{{\cal T}}_{contact} \; .
\ee

This simple example illustrates not only that, as 
expected~\cite{Coleman,Coleman_book}, total
{\it on-shell} amplitudes for a given process are invariant under field 
redefinitions, but it also shows the interplay between ``off-shell''
effects from vertices and contact terms. This makes it impossible to define
``off-shell'' effects in a unique, representation independent fashion.

Our considerations above concerned only rather simple vertices at
the tree level. The close connection between off-shell effects in a vertex and
contact terms also exists when we consider dressed vertices, as will be shown
at the one-loop level (see Appendix B). 
It can be made plausible with the following example that concerns the 
dependence of the vertex on the invariant mass $p^2$.
Consider, for simplicity, a scalar vertex for an initially on-shell 
particle together with the subsequent propagation. 
By expanding the vertex around the on-shell point,
\be\label{hur}
{ \Gamma(q^2, m^2, p^2)\over p^2-m^2 } = {\Gamma(q^2, m^2, m^2)\over p^2-m^2} 
+ {\partial\Gamma\over \partial p^2}(m^2) + \cdots \; ,
\ee
one finds that the propagator gets cancelled in the second and
higher order terms. Thus, off-shell 
effects in the pole terms through the dependence of the vertex on 
the scalar variable, $p^2$, can
also be related to contact terms. Equations (\ref{nea}) and (\ref{n2}) 
are a specific example of this. 
The above seems to suggest that it is possible to find a representation 
for an amplitude where $K(q^2,w)$ has no off-shell dependence, {\it i.e.} no 
dependence on $w$, by keeping enough terms in Eq. (\ref{hur}) and 
introducing the corresponding contact terms. However, 
the Taylor expansion implicit in Eq. (\ref{hur}) is valid only up to the 
first branch cut, {\it i.e.}
the pion threshold. Thus, this procedure is, for example, not valid in 
calculations of pion electroproduction on a nucleon.
In Compton scattering below the 
pion threshold the shifting of the dependence 
on the nucleon invariant mass to contact terms is possible.

The above example also showed how the transformation in Eq.\ (\ref{Dyson})
adds one
power of the nucleon four-momentum to the asymptotic behavior of the 
original vertex, Eq.\ (\ref{tria3}). Two powers can be added by 
considering a transformation involving derivatives, such as
\be\label{deriv1}
   \psi \rightarrow \exp\Bigl[ \beta \g5 \tp \f \Bigr] \psi \ .
\ee
To leading order in $\beta$, this transformation
generates the following $\beta$-dependent interaction terms
\bea\label{deriv2}
L^{[\beta]} &=& -2m\beta \bar\psi \g5 (\tp \f) \psi
                + i\beta \bar\psi \g5 (\tp \f)(\tp \psi)\nnn\\
            &&  - i\beta (\bar\psi \tp) \g5 (\tp \f) \psi 
                - i\beta \bar\psi \g5 (\partial^2 \f)\psi\nnn\\
            &&  - i\beta \bar\psi \g5 (\partial\f)(\partial \psi) 
                - i\beta (\bar\psi \partial) \g5 (\partial\f) \psi \ .
\eea
We readily obtain for the contribution of the $\beta$-dependent terms to the 
half-off-shell vertex
\be\label{deriv3}
   \Gamma_{5}^{\beta}(p',p) u(p) = \beta(m^2-p'^2)\g5 u(p) \ ,
\ee
which vanishes on-shell, as anticipated. Higher powers in the nucleon momenta
can be obtained by using transformations involving higher derivatives.
Of course, one can perform transformations acting on the nucleon 
field that induce not just a $p'^2$ dependence,
but also a combined $p'^2$ and $q^2$
dependence of the half-off-shell vertex, $\Gamma^5$. For example, the 
transformation $\psi$ 
$\rightarrow$ $\exp(i\beta\g5\partial^2\phi)\psi$ induces a new term
$\beta (\thru p'-m)q^2\g5 u(p)$. 

Observations similar to those we made for the strong form factor can also be
made for the electromagnetic vertex in QED by starting with the
QED Lagrangian and transforming the electron field. 
The electromagnetic vertex obtained at tree level from the QED Lagrangian 
is simply $-i e \gamma^\mu$ for on and off-shell electrons.
Applying the transformation $\psi$ $\rightarrow$ 
$ \exp(\beta\partial^2\!{\thru A})\psi$ changes, for example, the 
half-off-shell vertex to
$ -i[e+\beta q^2(\thru p'-m) \gamma^\mu]u(p)$. The $\beta$-dependent 
part of this vertex vanishes on-shell, as expected. 

\section{sidewise dispersion relations}

We now turn to the method of sidewise  dispersion relations which seems 
to suggest that one can uniquely obtain the off-shell
form factor from experimentally measurable phase shifts. 
There are two main issues we would like to address here. The first is
where does the repesentation dependence discussed in the previous section
enter the sidewise dispersion relation method. The second is
the validity of certain approximations, related to the treatment of
inelastic channels, that have been used in the literature.
Dispersion relations are expressions relating the real part of a 
function, such as a Green's function, to a principal value integral over 
its imaginary part. Physically, the requirement of causality implies 
the analyticity properties of such functions~\cite{BJ} which 
allows one to obtain dispersion relations.
%%% In field theoretical applications one usually takes as the 
%%% ``dispersion variable'' of these functions one 
%%% of the kinematical variables. 
Scattering amplitudes, for example, are real analytic functions of the 
energy E when regarded as a complex variable, {\it i.e.} $f(E)=f^*(E^*)$.
In the case of form factors of a particle, usually the four momentum transfer 
to the particle is used as the dispersion variable.

As shown by Bincer~\cite{Bincer} using the reduction formalism~\cite{LSZ},
one may analytically continue both the electromagnetic and the strong nucleon 
form factor not only as a function of the momentum transfer, but 
also of the invariant mass of the off-shell nucleon.
For the function $K(q^2, w)$ 
(we henceforth denote the dispersion variable by $w$)
appearing in the half off-shell
strong vertex, Eq.\ (\ref{kg}), he showed 
that it is a real analytic function of $w$ with cuts 
along the real axis starting at $w$ = $\pm(m+\mu_{\pi})$ and extending to 
$\pm \infty$. Furthermore, $K(q^2, w)$ is purely 
real along the real axis in 
the interval $-(m+\mu_{\pi})$ $<$ $w$ $<$ $(m+\mu_{\pi})$. Thus, $K(q^2, w)$ 
satisfies dispersion relations, termed ``sidewise'' to emphasize that the 
dispersion variable is now the nucleon four momentum, $w=\sqrt{p'{}^2}$.
Using Cauchy's theorem, one obtains
\be\label{B_strong}
    \Re K(q^2, w)={1\over \pi}{\cal P}\int_{m+\mu_\pi}^{\infty} d w'
    \Biggl[ {\Im K(q^2,w')\over w'-w} + {\Im K(q^2,-w')\over w'+w} \Biggr] \ ,
\ee
provided that $|K(q^2, w)|$ vanishes fast enough for
$|w|  \rightarrow \infty $. If, for example, 
$|K(q^2, w)|$ approaches a constant as $|w|$ $\rightarrow$ $\infty$, 
one must consider a once-subtracted dispersion relation for $K(q^2, w)$;
\bea\label{sub}
    \Re K(q^2,w) &=& \Re K(q^2,w_0)+ {(w-w_0)\over \pi}{\cal P}
    \int_{m+\mu_\pi}^{\infty}dw' \nnn\\
    \Biggl[\!\!\!\!\!&&\!\! {\Im K(q^2,w')\over(w'-w_0)(w'-w)} - 
    {\Im K(q^2,-w')\over(w'+w_0)(w'+w)}
    \Biggr] \ ,
\eea
where $w_0$ is a ``subtraction point'', most conveniently
taken to be the nucleon mass, $w_0=m$, where $K(q^2, m)$ is 
(experimentally) known. Evidently, 
if $|K(q^2,w)|$ grows like $|w|^n$ ($n$ an integer) as 
$w$ $\rightarrow$ $\infty$, 
$n$+1 subtractions must be performed which introduce the same number of
{\it a priori} unknown subtraction constants into the dispersion relation.
The role of subtractions in the sidewise dispersion method is important.
Since we only know $K(q^2,w)$ at the on-shell point,
$w=m$, a need for more than one subtraction will spoil any possible 
predictive power. In cases where the vertex function 
is not known at the on-shell point, as, {\it e.g.} in
the electromagnetic vertex of the nucleon, even one subtraction
will destroy predictive power.

For our discussion below we are interested in the case where
the pion is on its mass shell, {\it i.e.} in $K(m_{\pi}^2,w)$. 
It is useful to note that starting from
Eq.~(\ref{sub}) one can obtain~\cite{Bincer}
\bea\label{phase1}
    |K(m_{\pi}^2,w)| &=& |K(m_{\pi}^2,m)| \exp\Biggl\{
    {(w-m)\over \pi}{\cal P}\int_{m+\mu_\pi}^{\infty}
    dw' \nnn\\&&\biggl[ {\phi(w')\over(w'-m)(w'-w)} 
    - {\phi(-w')\over(w'+m)(w'+w)} \biggr] \Biggr\} \ ,
\eea
where $\phi (\pm w)$ is the phase of $K(m_{\pi}^2,w)$ along the positive (+)
or negative (-) cut. Thus, $K(m_{\pi}^2,w)$ can be determined if these phases
are known. 

So far, dispersion relations just reflect analyticity properties of 
Greens functions and are void of any predictive power. This changes
when one makes use of unitarity constraints that provide additional
relations between the real and imaginary parts of the Green's function.
The simplest example of the power 
of using analyticity and unitarity in conjunction is the 
forward amplitude for the scattering of light with 
frequency $\omega$ from atoms. Unitarity implies that the forward
amplitude for positive frequencies is related to 
the total cross section through the optical theorem, 
\be\label{optical}
   \Im f(\omega) = {\omega \over 4\pi} 
   \sigma_{\mbox{tot}}(\omega)\ , \; \omega > 0 \ ,
\ee
leading, with one subtraction, to the famous Kramers-K\"onig relation
\be\label{KK}
   \Re f(\omega) = \Re f(0) + {\omega^2\over 2\pi^2}{\cal P}\int_{0}^{\infty}
   dw'{ \sigma_{\mbox{tot}}(\omega')\over (\omega '^2-\omega ^2)} \ ,
\ee
which allows the determination of $f(\omega)$ from the experimentally measured
total cross section. 

For the the meson-nucleon T-matrix, unitarity implies the 
well-known matrix equation
\be\label{ena}
   \Im{T} = T T^\dagger \ ,
\ee 
from which the optical theorem follows. 
This equation assumes a simple form after projecting onto 
states of total angular momentum, J, parity, P, and 
isospin, T. We will consider in the following sections a simple situation 
where there are two reaction channels, $\pi N$ and $\eta N$. The 
T-matrix may then be written in the general form
\be\label{param}
   T^l = \xara(\tpp,\tpe,\tpe,\tee) \ ,
\ee 
where $l$ labels the quantum numbers {J,P,T}, and the two-body channels are
denoted by $\pi$ and $\eta$. Furthermore, $\delta^{l}_{\pi}$ and
$\delta_{\eta}^{l}$ are the elastic scattering phase
shifts for $\pi N$ and $\eta N$ scattering, respectively, given by  
\be\label{tan2d}
   \tan 2\delta^l_i ={2\Re T^l_{ii}\over 1 -2\Im T^l_{ii}},
   \quad i=\pi,\eta \ ,
\ee 
and $\rho_{l}$ is the corresponding inelasticity parameter.
The T-matrix is symmetric since time-reversal invariance has been assumed. 
Below the $\eta$ threshold, only $T^{l}_{\pi\pi}$ is nonzero, and the
familiar elastic form of $T$ is obtained,
\be
   T^l_{\pi\pi} = \sin \delta_l^{\pi} e^{i\delta^l_{\pi}} \; .
\ee

For $w$ $>$ 0, $T_{\pi\pi}$ describes 
$\pi$N scattering in the $P_{11}$  partial wave 
 ($l$ = {$1/2 ^{+},1/2$}) and for $w$ $<$ 0 scattering in the $S_{11}$ 
partial wave ($l$ = {$1/2 ^{-},1/2$}). 
The consequences of unitarity for $K(m_{\pi}^2,w)$ may be obtained by 
looking at its absorptive part which receives contributions from
{\it physical} on-shell intermediate states. Unitarity provides for K,
which is now a vector in the space of the different reaction 
channels, the constraint~\cite{Bincer,Bos,Epstein}
\be\label{imktk}
\ik = F^{-1} T F K^* \ .
\ee
Here $\finv$ and $F$ are phase space factors (see appendix A). 
For the $\pi NN$ form factor, this constraint can be written as
\bea\label{aa1}
    \Im K_{\pi}(m_{\pi}^2,w) &=& \theta (|w| -m-\mu_{\pi})
    T_{\pi\pi} (w) K_{\pi}^* (m_{\pi}^2,w)\nnn\\
    & +& \theta (|w|-w_{T}) A(w) \ ,
\eea
where $w_T$ is the threshold energy of the first inleastic channel.
The first term on the right-hand-side of Eq.\ (\ref{aa1}) arises from the 
intermediate pion-nucleon two-body state and the second term represents 
contributions from intermediate states with higher masses, {\it e.g.}
$\pi \pi N$, $\eta N$, $K \Lambda$, etc. 

For $w < w_T$,  the last term in Eq.~(\ref{aa1}) does not contribute
and one sees from this equation that the phase of the form 
factor for the $\pi NN$ vertex, $\phi_\pi$ = \Arg$(K_\pi)$, 
is determined { by} the elastic 
$\pi N$ phase shift, defined in Eq.~(\ref{tan2d}),
\be\label{a2} 
   \phi_{\pi} = \delta^l_{\pi} = {1\over 2}\tan^{-1}\left( 
   2 \rt_{\pi\pi} \over 1 - 2\imt_{\pi\pi} \right) \  .
\ee
We note that
since the phase shift is a representation independent observable quantity, 
both the real and imaginary parts of the representation dependent 
off-shell form factor, Re$K_{\pi}$
and Im$K_{\pi}$, must change under a field transformation such that
the phase, $\phi_\pi$ $\equiv$ $\arctan( \Im K_\pi / \Re K_\pi )$, 
remains unchanged for $w < w_T$.

The use of the dispersion technique to obtain the vertex function, 
$K(w)$, for $w \ne m$ with only experimental input
faces problems in practice. 
In order to obtain the off-shell form factor from 
Eq. (\ref{phase1}), the phase must be 
known up to infinite energies. Therefore, 
it is clear that one must make approximations about the behavior of
the elastic phase shift for high energies and also about the 
contributions coming from the inelastic channels for $w > w_T$. 
Two such approximations have been proposed in the literature. 

The simplest assumption is to ignore inelastic contributions, 
{\it i.e.} set $ A = 0$ in Eq.\ (\ref{aa1}), which would
be justifiable if the dispersion integral is dominated by the
interval where the contribution from $A$ is small compared to the elastic term.
This is referred to as the ``threshold'' approximation \cite{Bincer}.
It amounts to assuming that Eq.~(\ref{a2}) remains valid for all energies and
allows one to evaluate $K(w)$ in terms of the elastic phase shift.
As shown in Ref.~\cite{Epstein}, the threshold assumption
implies $\rho_{l}^2$ = 1, which is quite unrealistic as soon as one 
gets above the threshold for the $\pi \pi N$ channel. 

To avoid this problem,
Epstein adopted in the dispersion analysis of 
the off-shell $\pi N N$ form factor a suggestion by Goldberger and 
Treiman~\cite{Treiman} which leads to a different {\it ad hoc} 
prescription to deal with the dispersion integral.
Consider the right hand side of Eq.~(\ref{aa1}), which, 
although it involves complex quantities, must nevertheless 
be real. This leads to the following conditions for the combined effect of 
the inelastic states contained in the complex quantity $A$,
\bea\label{Aconstraint}
	\Re A &=& \Im K_{\pi} -  \Re T_{\pi\pi} \Re K_{\pi} - 
		\Im T_{\pi\pi} \Im K_{\pi} \nonumber \\
	\Im A &=& \Re T_{\pi\pi} \Im K_{\pi}  -  
                    \Im T_{\pi\pi} \Re K_{\pi}  \ .
\eea
Epstein assumed that the inelastic
channels will not generate a significant real part for $A$, 
{\it i.e.} $\Re A=0$. This leads to a different expression 
for the phase $\phi_{\pi}$ of the $K_\pi$ form factor in terms 
of the elastic $\pi N$ T-matrix~\cite{Epstein},
\bea\label{aa2}
    \phi_{\pi} &=&\tan^{-1}\left( 
    \rt_{\pi\pi} \over 1 - \imt_{\pi\pi} \right)\nnn\\
    &=& {1\over 2}\tan^{-1}\left( 
    2 \rt_{\pi\pi}(1 - \imt_{\pi\pi}) \over (1 - \imt_{\pi\pi})^2 -
    (\rt_{\pi\pi})^2 \right) \  \  .
\eea
In the rest of this paper, we will refer to this approximation for simplicity as
the
``Goldberger-Treiman'' approximation. Of course, 
by setting $\Im A=0$ as well, from Eq.~(\ref{Aconstraint}) 
we would then again obtain $\f_{\pi} = \delta^l_{\pi}$, 
the threshold approximation. Below the inelastic threshold, the unitarity 
constraint, Eq.~(\ref{ena}), reads $(\imt_{\pi\pi})^2+(\rt_{\pi\pi})^2=
\imt_{\pi\pi}$, and simple inspection shows that that (\ref{a2}) and
(\ref{aa2}) agree, but above the inelastic threshold they can be
quite different, especially if a resonance is present (see section IV).
The general problem of the choice of the phase above the inelastic threshold 
was also discussed in connection with the dispersive analyses of the pion 
{\it elastic} electromagnetic form factor \cite{BJ}.

Even knowing all the relevant T-matrix elements, it is not at all 
straightforward to solve for $\phi_{\pi}$. To illustrate this,
we stay with the case when there are only two channels present, the two-body 
states $\pi N$ and $\eta N$. Defining $K(\pm w)$ $\equiv$
$K(m_{\pi}^{2},\pm w)$,
we find from Eq.\ (\ref{imktk}) that
\bea\label{kun}
    \Im K_{\pi}(+w) &=& T_{\pi\pi}^{P11}(+w)   K_{\pi}^*(+w) +
    {F_{\eta\eta}^{-} \over F_{\pi\pi}^{-}}
    T_{\pi\eta}^{P11}(+w)K_{\eta}^*(+w) \nonumber \\
    \Im K_{\eta}(+w)&=& T_{\eta\eta}^{P11}(+w)K_{\eta}^*(+w) +
    {F_{\pi\pi}^{-} \over F_{\eta\eta}^{-}}
    T_{\pi\eta}^{P11}(+w) K_{\pi}^*(+w) 	         \ ,
\eea
and similar equations for $K(-w)$. 
In the two channel case, the term $A$ in Eq. (\ref{aa1}) is
\be\label{aeta}
   A (\pm w) = T_{\pi\eta} ( \pm w) K_{\eta}^* ( \pm w) 
   \sqrt{ { q_{\eta}(E_{\eta} \mp m) \over q_{\pi}(E_{\pi}
   \mp m) } } \; ,
\ee
where $q_{\pi(\eta)}$ is the $\pi$($\eta$) three-momentum in the cm 
frame and $E_{\pi(\eta)}$ = $\sqrt{q_{\pi(\eta)}^2 +m^2}$. As for
$T_{\pi\pi}$, for $w$ $>$ 0 $T_{\eta\pi}$ is given by the $J^{P}T$ = 
${1/2}^{+},1/2$ partial wave while for $w$ $<$ 0 it is given by the 
${1/2}^{-},1/2$ partial wave.

Equations (\ref{kun}) seem to provide the desired constraints
for extracting the phases $\phi_{\pi}$ = Arg$(K_{\pi})$ 
and $\phi_{\eta}$ = Arg$(K_{\eta})$ from the meson-nucleon T-matrix
without resorting to any of the aforementioned approximations.
First, one eliminates the magnitudes $|\G5_{\pi}|$
and $|\G5_{\eta}|$ from Eq. (\ref{kun}). Since the phase space factors 
cancel out, one obtains
\be\label{a11}
T_{\pi\eta}^2e^{-i(\phi_{\pi}+\phi_{\eta})}=
              \Bigl[\sin\phi_{\pi}-T_{\pi\pi}e^{-i\phi_{\pi}}\Bigr]
              \Bigl[\sin\phi_{\eta}-T_{\eta\eta}e^{-i\phi_{\eta}}\Bigr] \ .
\ee
The real and imaginary parts of Eq.~(\ref{a11}) provide two equations that 
should allow the determination of $\phi_{\pi}$ and
$\phi_{\eta}$ from the T-matrix elements. 
However, this is not possible because the two resulting equations are in fact 
identical due to the unitarity constraint for the T-matrix. To see this, notice 
that $\ik = \finv T F K^*$ implies both
\be\label{impose1}
   \ik = \finv (\rt) F (\rk) + \finv (\imt) F (\ik) 
\ee
    and
\be\label{impose1b}
   (\imt) F (\rk) = (\rt) F (\ik) \ .
\ee
Using Eq.  (\ref{impose1b}), Eq.\ (\ref{impose1}) reads
\bea
    \finv \Bigl[ 1-\imt \Bigr] F (\ik) &=& \finv (\rt) F (\rk) \nnn\\
    \Rightarrow   \Bigl[ 1-\imt \Bigr] \rt^{-1} \imt &=& \rt 
    \label{impose2}\ .
\eea
Using the real part of the unitarity condition for the T-matrix, 
$T T^\dagger=\imt$, yields $(\rt)^2+(\imt)^2=\imt$, while the imaginary part
results in the vanishing of the commutator $[\rt,\imt]=0$. This 
shows that Eq.~(\ref{impose2}) is just the { T-matrix} unitarity constraint
and the real and imaginary parts of  Eq.~(\ref{impose1}) do not
provide independent equations allowing the determination of the
phase of K from the on-shell T-matrix. 

This does not necessarily imply that sidewise dispersion 
relations cannot be used to determine $K$ ({ provided that one 
subtraction is enough}), but more work needs to be 
done. Above the eta threshold we can use Eq.\ (\ref{kun}) 
to determine $\Im K$ in terms of $\Re K$ and the on-shell T-matrix. 
This may then be substituted into Eq.\ (\ref{sub}) to obtain a coupled 
set of Fredholm-like integral equations for $\Re K$. The problem, as 
shown below, is that between the pion and eta thresholds 
$\Im K_{\eta}$ is expressed in terms of an off-shell T-matrix 
element; it might then be possible through dispersion relations to 
determine $T^{\pi\eta}$ at the needed off-shell points in terms of
on-shell information. A more detailed investigation of this possibility
is beyond the scope of this paper.

In the (hypothetical) case of a single channel system it seems
to be possible to determine the phase $\phi$ and thus also the function 
$K(w)$ for the off-shell vertex in a model independent fashion using 
the observable phase, $\delta$, of the on-shell T-matrix.
This appears to be in contradiction with the observation in section II 
that the off-shell form factor changes when we carry out field 
transformations. How can this be reconciled with the sidewise dispersion 
relations that express $K(w)$ in terms of observable quantities? 

The  answer lies in the fact that in the sidewise dispersion relation approach
the number of necessary subtractions is {\it a priori} unknown. Indeed, 
different choices of the nucleon interpolating field will in general lead 
to different asymptotic behaviors of the off-shell form factor. The
examples given in Sect.~II illustrate this point. From  
Eq.~(\ref{tria3}) we see that $K(w) =g$, {\it i.e.}
is of order ${\cal O}(1)$ as $w \rightarrow \infty$.  On the other hand, 
the vertex function, Eq.\ (\ref{kw2}), obtained from the transformed 
Lagrangian is of order ${\cal O}(w)$ at infinity. Thus, 
the ``representation dependence'' in 
sidewise dispersion relations shows up in the {\it a priori} unknown needed 
number of subtractions.  As previously remarked, any predictive power of the 
sidewise dispersion relations method will be lost if two (or more) 
subtractions are necessary since we only know the form factor at the
physical point $w=m$. Another way to improve the convergence of the
dispersion integral is to consider the derivative of $K(w)$.
However, none of its derivatives with respect to $w$ at $w = m$ are known
and therefore no information about an off-shell point can be obtained.

As dispersion relations do not depend on a particular Lagrangian, it is useful 
to look at the above discussion for the vertex function $K$ in a different way
and to contrast it with the dispersion relations for the pion--nucleon
scattering amplitude. Consider the unitarity constraint, 
Eq.~(\ref{imktk}): evidently, it remains valid under the 
replacement $K$ $\rightarrow$ $f(w)K$, 
where $f(w)$ is a real function of $w$, reflecting a different off-shell 
behavior. If $f(w)$ is a polynomial in $w$ and one has $f(m)$ =1, then the 
analytic properties of $K$ are not changed and $K$ still satisfies a 
dispersion relation. However, in general (additional) 
subtractions will be needed, and these subtractions have to be done at an 
unphysical points, $w \neq m$, and therefore cannot be done model 
independently. 
When using the dispersion relation approach for the T-matrix, we also
may need subtractions to make the integrals converge. However, for
this purpose we can do these subtractions at different energies where we 
have experimental information about the T-matrix. In some cases, this makes it 
possible to determine the pion-nucleon T-matrix at an 
unphysical point through dispersion relations,
while the off-shell {\it form factors} can never be uniquely determined. 
Notice also that our discussion does not imply that dispersion relations 
for the electromagnetic form factor, with the {\it momentum transfer} $q^2$ 
as the dispersion variable, show any representation dependence. In
this case the form factor $F(q^2)$ can be measured for a 
number of values of the four-momentum transfer,~$q^2$.

\section{A coupled-channel unitary model}

In the previous section we have discussed two inherent difficulties of 
the sidewise dispersion relation approach applied to the 
off-shell form factors. The first one is the {\it a priori} unknown
number of subtractions, which reflects the ``representation dependence''. 
The second difficulty is related to determining $\phi$, the phase of the 
vertex function, in terms of observable physical 
quantities: one is in general unable to properly take into account the 
contribution of all possible intermediate states to the absorptive 
(imaginary) part of the form factor. With respect to this second difficulty,
two approximations had been proposed in the literature; the threshold
approximation, \equ(a2), and the ``Goldberger-Treiman'' \cite{Epstein}
approximation, \equ(aa2). 
In this section we study these approximations by using a model with a 
nucleon interpolating field that leads to a $K$ satisfying
a once-subtracted dispersion relation.
Even after assuming the validity of only one subtraction, a precise
determination of $K$ through
the dispersion relations remains extremely difficult (if not 
impossible). It is therefore interesting to see in the framework of a simple 
model under what circumstances the two approximations to the phase discussed 
in the previous section can be trusted to give reasonable results for $K(w)$. 
Indeed, $K(w)$ has already been ``extracted'' from the $\pi N$ phase shifts
using the Goldberger-Treiman approximation \cite{Epstein}, but the
approximation itself has not been examined. Since we will use a meson
loop model, this allows us also to examine, {\it e.g.} the behavior of 
the absorptive part 
of the strong form factor under field redefinitions, extending the studies 
of section III beyond the tree level.

We construct a unitary T-matrix based on the toy model meson-nucleon 
Lagrangian~\cite{others}
\bea\label{tria}
   L &=& {1\over 2}\Bigl( \dm\F\dM\F - \F\mmm^2\F\Bigr) + \bar\psi(i
   \thru{\partial}- m)\psi \nnn\\
   && - i\bar\psi\g5 G\F\psi + 
   {1\over 2}\bar\psi\F\L\F\psi \ .
\eea
Apart from the conventional fermionic part describing the nucleon
with mass $m$,  we take into account two isoscalar mesons, given by
the two-component field $\F$, and their mass matrix, $\mmm$,
\be\label{t1}
   \F=\lll(\pi,\eta)  \>,\qquad\qquad
    \mmm=\xara(\mu_{\pi},0 , 0 ,\mu_{\eta}) \ ,
\ee
where we use the suggestive names ``pion'' and ``eta''. 
We assume a pseudoscalar three-point meson-nucleon 
coupling, $G$, and a scalar four-point meson-meson-nucleon 
coupling, $\L$, where
\be\label{t2}
   G=\lll(g_{\pi},g_{\eta}) \>,\qquad\qquad
   \L=\xara( {\lpp},{\lpe},{\lep},{\lee} ) \ .
\ee
The $\L$ matrix has the dimension of an inverse mass, while $G$ is 
dimensionless. The non-vanishing off-diagonal elements,  
$\lpe$ and $\lep$, couple the two meson channels. For simplicity, we
make the choice $\lep=\lpe=\sqrt{\lpp\lee}$ (see Appendix A).

While we cannot solve this model exactly, it is possible
to select an infinite subset of diagrams which satisfies the necessary
analyticity and unitary properties. We do that by treating the 
three-point ``$G$'' coupling to leading order only, while summing 
higher order contributions generated by the ``$\Lambda$'' interaction. 
Only two-particle intermediate states, {\it i.e.} $\pi N$ and $\eta N$, but 
not $\pi\pi N$ or $\eta \pi N$ are considered. Our approach does not 
satisfy crossing symmetry and, moreover, there are no meson loops that 
connect the incoming and outgoing nucleons; they would be either of
second order in $G$, or have three-particle intermediate states. This  
selection of contributing diagrams does not generate a $q^2$ dependence for 
the form factor (in other words, what one usually refers to as the ``on-shell''
form factor is trivial in this model). However, it {\it does} 
generate a { nontrivial dependence on the invariant mass $p'^2$ of the 
off-shell nucleon and satisfies two-body-unitarity.
We should also emphasize here that the truncation to ``two-body-unitarity'' 
is an approximation, but it is the validity of 
the approximations made {\it on top} of our assumptions that we wish to 
test here.  

The diagrams that can contribute to meson-nucleon 
scattering with our restrictions are shown in Fig.\ \ref{tmatr},
and those contributing to the half-off-shell meson-nucelon form 
factor in Fig.\ \ref{kstrong}. The 
external and internal mesons may be either pions or etas. As the
``$\Lambda$'' interaction is separable, we can express the geometric 
series for the T-matrix in a closed form
\bea 
    \T  &=& \L + \L\I\L + \L\I\L\I\L + \dots \nonumber\\
    &=&   (1 -\L\I )^{-1}\L \label{tmatrix} \; ,
\eea
where $\I$ = diag$(\I_\pi,\I_\eta)$ is a diagonal matrix whose ($\pi\pi$) 
and ($\eta\eta$) entries are pion-nucleon and eta-nucleon
loop integrals, respectively.
The interested reader is referred to Appendix A for details of 
results stated without proof throughout this section.
The integrals $\I$, and therefore the $\T$ matrix,
depend only on the total center of mass momentum 
squared, $s = w^2 = (p+q)^2$, and have no angular dependence.
Using the projection operators defined in Eq. (\ref{prop})
and the standard partial wave projections, it is easy to show that
$P_{+}$ projects only into the $f_{0+}$ partial wave and 
$P_{-}$ only into the $f_{1-}$ partial wave, that is
\be\label{tmat1a}
\T=\T^{0+}P_{+} + \T^{1-} P_{-} \; .
\ee
Taking into account the appropriate phase space factors $F$, it can
be shown (see Appendix A) that $T=F \T F$, 
satisfies unitarity for $|w|<m+M $, where $M$ is the cutoff needed to 
regularize the loop integrals $\I$,
\be\label{ale2}
 \Im(T^{S11}) =  T^{S11} (T^{S11}{})^\dagger\label{exi2} \ ,
\ee
and analogously for $T^{P11}$.
Thus, the $T^{l}_{ij}$ may be written in the form given in Eq.\ (\ref{param})
where $\dpl$ ($\del$) are the pion (eta) phase shifts (S-wave for $l=0$
and P-wave for $l=1$) and $\rho_l$ are the corresponding inelasticities
($\rho_l=1$ below the eta threshold, $w_T=m+\mu_{\eta}$).
For a numerical study of these form factors 
we take for $m$ and $\mu_{\pi}$ 0.939 GeV and 0.14 GeV, respectively, and 
choose $\mu_{\eta}$ to be 0.42 GeV, since at $w$ = 1.36 GeV = $m+\mu_{\eta}$ 
the $P_{11}$ inelasticity starts to deviate from unity. The $\L$ couplings are 
chosen to reproduce some qualitative features of the physical pion-nucleon 
scattering phase shifts and inelasticity, in particular, a resonance appearing
above the inelastic threshold. While the actual $\pi N$ scattering
amplitude exhibits this feature in
both the $P_{11}$ and $S_{11}$ channels, our model is 
too simple to simultaneously produce resonances in both channels. We therefore 
concentrate on the $P_{11}$ channel, and show in Fig.\ \ref{phsh} the phase 
shifts and inelasticity parameter obtained with 
$\lambda_{\pi\pi}$   = 0.5 MeV$^{-1}$ and $\lambda_{\eta\eta}$ = 
0.8 MeV$^{-1}$, which
leads to a resonance in the $P_{11}$ channel with a substantial inelasticity.
This parametrization yields $K_{\eta}(m)/K_{\pi}(m) = - 0.87$.
As mentioned, the presence of a finite 
cutoff violates unitarity for $w$ $>$ $m+M$, and we therefore
use a large cutoff, $M$ = 10 GeV. 

The  half-off-shell strong vertex is in our model
generated by the series in Fig.\ \ref{kstrong}, and gives
\bea\label{strongf}
   [\Gamma^5(p',p)]^T u (p) &=& 
			\g5 G^T \left[1-\I(w)\L\right]^{-1}  u (p) \nnn\\ 
\Rightarrow 
   \Gamma^5(p',p) u (p) &=& 
			\g5 \left[1-\L\I(w)\right]^{-1}G  u (p) \ .
\eea
Here the transpose acts on the channel space 
indices only, and $w=\sqrt{(p+q)^2}$. With our assumptions, 
the pion-nucleon strong vertex function reads
\be\label{gpnn}
   K_{\pi}(w) = {(1-\lee\Ip(w) )\gp + \lpe\Ie(w) \get\over
                       1-\lpp\Ip(w) -\lee\Ie(w) } \; ,
\ee
where we have ignored graphs of order ${\cal O} (G^3)$ and higher, as
well as intermediate states with more than two particles. Meson loops on 
the on-shell nucleon need not to be considered since their contributions 
can be absorbed in the definition of the on-shell vertex. The vertex function 
$K(w)$ is determined through the P-wave on-shell scattering amplitude, 
$T^{P11}(w)$, and $K(-w)$ by the S-wave on-shell amplitude, $T^{S11}(w)$. 
In the examples below, the values of $g_{\pi}$ and $g_{\eta}$ will be varied 
to change the ratio of the on-shell form factors, $K_{\eta}(m)/K_{\pi}(m)$.

As shown in appendix A, the unitarity equation for $K$, Eq.\ (\ref{kun}), is 
satisfied in this model below the cutoff. As mentioned above, below the pion 
threshold, $|w|$ $<$ $m+\mu_{\pi}$, one has $\Im K$ = 0. Between the pion 
and eta thresholds, $m+\mu_{\pi}$ $<$ $|w|$ $<$ $m+\mu_{\eta}$ we have 
\bea\label{kun2}
\Im K_{\pi}(+w)  &=& T_{\pi\pi}^{P11} K_{\pi}^{*} (+w) \nonumber \\
\Im K_{\eta}(+w) &=&  (F_{\pi\pi}^{-})^2 \T_{\pi\eta}^{P11}
K_{\eta}^{*}(+w) \ ,
\eea
where $\T_{\pi\eta}^{P11}$ is the $\pi\eta$ matrix element multiplying
the $P_{-}$ operator in Eqs. (\ref{tmat1a}) and (\ref{tmat1}), in this case 
evaluated at an off-shell point. Although the form of Eq. (\ref{kun2}) is 
specific to our model, it is true in general that Im$K_{\eta}$ is non-zero 
between the thresholds and in this region
is related to off-shell quantities.

Let us now discuss the dispersion relations for the off-shell form factors
in this model. While the presence of the cutoff violates unitarity, it does 
not affect the validity of the dispersion relations.
Due to the choice of a large cutoff the dispersion integral has largely 
converged by the time the cutoff is reached. It is well-known that the functions
$\I_{\pi}$ and
$\I_{\eta}$ satisfy once-subtracted dispersion relations \cite{IZ}. 
$K_{\pi}$ and $K_{\eta}$ have the same analytical structure as 
$\I_{\pi}$ and $\I_{\eta}$, apart from possible additional
poles on the first Riemann sheet. Furthermore,
the number of needed subtractions may be different, but it is easy to
to establish in our model that $K$ also satisfies a once-subtracted dispersion 
relation. On the other hand, the existence of poles in the complex $w$ plane 
is more difficult to assess.
We have simply established their absence numerically
by showing that the once-subtracted dispersion relation is satisfied to six 
significant figures for -2 GeV $<$ $w$ $<$ 2 GeV. 

In the analysis of the pion-nucleon vertex function by Epstein~\cite{Epstein} 
and by J.~W.~Bos \cite{Bos}, it was found that the Goldberger-Treiman 
approximation leads to a much smoother off-shell behavior of $K$ than in
the threshold approximation. This can be easily explained: if one uses 
the threshold approximation, the phase of the form factor is {\it given} by 
the scattering phase shift and we therefore expect the function K to
show resonance behavior. When the $\pi NN$ phase shift passes through $\pi$/2 
the threshold approximation to $\Re{K}$ will
change sign and $\Im{K}$ peaks. On the other hand,
using the Goldberger-Treiman assumption, $\phi_\pi$ is constrained to
be in the interval $-\pi /2$ $<$ $\phi_\pi$ $<$ $\pi /2$. This is easily seen 
from  Eq. (\ref{aa2}) with $\rho_l < 1$, which implies 
$T_{\pi\pi} < 1$. Therefore, $\Re{K}$ will not change sign 
since $\phi_\pi$ does not pass through $\pi$/2. Thus, we expect 
that the  Goldberger-Treiman approximation will generate a smooth
off-shell dependence, while the threshold assumption will generate an 
more rapid dependence on $w$ if there is a resonance in the 
scattering T-matrix, as happens in reality for $\pi N$ scattering 
as well as  with our $P_{11}$ phase shifts. 

Our model allows us to put these these previous analyses into perspective
and to confirm our qualitative expectations. In Fig.\ \ref{allap} we show 
the exact model results for $K_{\pi}$ with $g_{\pi}$ and $g_{\eta}$ 
adjusted to give $K_{\eta}(m)/K_{\pi}(m)$ =  $-1$ and 1, 
all other parameters as in Fig.\ \ref{phsh}. 
As expected, $K$ obtained from the threshold assumption
 (solid lines) displays a rapid $w$ variation
due the resonance in the $P_{11}$ channel (see Fig. \ref{phsh}),
while the Goldberger-Treiman approximation (dot-dashed lines) 
leads to a rather smooth energy dependence of $K$. 
Whether the Goldberger-Treiman or threshold approximation
is better cannot be answered in general. It depends on the details
of the dynamics. At $K_{\eta}(m)/K_{\pi}(m)$ =  $-1$, 
the  Goldberger-Treiman approximation seems to work well,
while at $+1$ it is the threshold approximation that works well.
We therefore conclude that neither approximation may be trusted
{\it a priori} at any $w$. Fig.\ \ref{allap} shows that there
can be large discrepancies between the exact model result and the
phase approximations even in the vicinity of the on-shell point.

As shown in Fig.\ \ref{phsh}, the inelasticity deviates 
signigicantly from unity for large values of $w$. 
That the qualitative features of the
two approximations discussed above are not due to this large 
inelasticity was confirmed by considering another parametrization 
(results not shown). A resonance in the $P_{11}$ channel 
can also be obtained with, {\it e.g.}
$\lambda_{\pi\pi}$ = 0.5  MeV$^{-1}$ and 
$\lambda_{\eta\eta}=0.05$ MeV$^{-1}$. Since $\lep$ = $\lpe$ = 
$\sqrt{\lpp\lee}$, this corresponds to a much weaker
coupling between the channels and the inelasticity remains 
close to one. The same features as in Fig. \ref{allap} were 
found, thus casting doubt on the use of the threshold approximation
in general. In fact, the threshold approximation requires 
$\rho_l$ = 1, while the Goldberger-Treiman is too restrictive
to allow for a resonant behavior of the form factor. Thus, 
neither of these approximations can be expected to be satisfactory.

The above observations are not based on some specific detail of our model,
but on rather general properties such as the existence of resonances in 
the T-matrix. The dependence on the details of the underlying reaction 
mechanism that we have shown with our simple model probably 
underestimates the real situation. In our model, the off-shell variation at
positive $w$ is mainly due to the resonance in the $P_{11}$ channel. For
example, the $S_{11}$ resonance in $\pi N$ scattering, absent in our model, 
would afflict the negative $w$ sector as well. 

It is also interesting to look at the model results for the eta form 
factor, $K_{\eta}(w)$. The results for the same set of 
couplings, $\L,$ $G$, as in Fig.\ \ref{phsh} are shown in 
Fig.\ \ref{keta}. Due to the simplicity of our ``toy model'', 
the results again only illustrate some general qualitative
features. At negative energies, we see the very pronounced 
effect due to the pion and eta thresholds. This effect is not 
visible for positive $w$ since the P-wave phase space 
suppresses the cusp. It can be seen that $K_{\eta}$ is complex 
even below the $\eta$ threshold and displays some rapid  
energy dependence around the $\eta$ threshold. These features 
arise due to the branch cuts associated with the thresholds, 
and therefore should be general features of the function 
$K_{\eta}$. The magnitude of these effects will of course depend 
on the model. Nevertheless, this casts doubt on the use of simple 
tree-level amplitudes with real coupling constants to extract 
the $\eta NN$ coupling constant, for example from photoproduction 
of etas~\cite{tiator}. 

\section{Summary and Conclusions}

Sidewise dispersion relations have been suggested in the literature as 
a method to obtain the electromagnetic and strong half-off-shell 
form factors of the nucleon. These form factors enter in calculations of,
{\it e.g.} nuclear processes or of two-step reactions on a free nucleon
where one includes the structure of the nucleon in terms of
dressed vertices.
We have focussed our discussion on the strong 
pion-nucleon vertex, where sidewise dispersion relations relate the 
half-off-shell strong form factor to on-shell 
meson-nucleon scattering. Two aspects of this approach were examined, 
its representation dependence and the validity of approximations that
have been used in the literature.

The strong vertex, or three point Green's function,
is not uniquely defined when one or both nucleons are not on their
mass shell. They are dependent on the representation one chooses for
the intermediate (off-shell) fields and therefore cannot be unambigously 
extracted from experimental data. In order to illustrate this representation 
dependence and to show how it enters in the sidewise 
dispersion relations, we used unitarily-equivalent Lagrangian models. 
Starting at the tree level, we showed how off-shell vertices do change under 
a change of representation, while the on-shell vertices
are oblivious to such changes. We then showed how in  on-shell amplitudes 
corresponding to two-step processes, this representation dependence of the
vertices manifests itself through contributions of pole as well as contact 
terms. This means that what one would call off-shell effects due to a vertex
in an amplitude
in one representation are related to contact 
terms in another. We then showed how the changes of representations 
can change the asymptotic behavior of the off-shell form factor,
thus requiring a representation-dependent number of subtractions in the
dispersion relation. In other words, representation dependence enters the
sidewise dispersion analysis through the number of necessary subtractions.
As the form factor is only known at the on-shell point, $w = m$, 
only one subtraction constant is known and the sidewise dispersion
 analysis thus has no predictive power for the vertex function.
We showed at the one-loop level in perturbation theory that not 
only the real, but also the ``absorptive'' imaginary part of the 
half-off-shell form factor, related to open physical channels, 
exhibits this representation-dependent asymptotic behavior.
 
Even when one chooses a particular representation ({\it i.e.} 
assumes one subtraction), one still faces problems when trying 
to obtain the corresponding off-shell vertex functions. 
These difficulties are due to the contribution of inelastic 
channels, {\it i.e.} other than $\pi N$ intermediate states. These channels 
contribute through the unitarity constraint that relates the half-off-shell 
vertex function to the meson-nucleon T-matrix, or scattering amplitude. 
Approximations how to deal with these channels in an {\it ad hoc} fashion
had been proposed in the literature, but their validity had not been examined.

In order to study these recipes, we introduced a very simple
coupled-channel, unitary model for the pion-nucleon system, 
where the inelastic channel is represented by an $\eta N$ 
intermediate state. We first established that the half-off-shell 
form factor in this toy model satisfies a once-subtracted 
sidewise dispersion relation and then compared 
this result to the results obtained from sidewise dispersion 
relations using these {\it ad hoc} prescriptions. We found that 
differences among the approximations and the exact model result
for the off-shell vertex functions can be sizable, particularly
when $w$ lies in the vicinity of resonances of the T-matrix, 
where the two prescriptions we tested produce very different 
results. We found that which of the two prescriptions is better, 
{\it i.e.} is closer to the exact model result, depends on 
details of the dynamics assumed in the model. Therefore, 
neither of the two approxiamtions is {\it a priori} preferred,
and the results one obtains by using such recipes remain questionable. 

We conclude that in practice sidewise dispersion 
relations cannot provide reliable and unique information about the structure of 
off-shell nucleons. The number of required subtractions is representation 
dependent and thus {\it a priori} unknown. Even if one chooses a particular
representation, the inclusion of the other reaction channels cannot be
dealt with without approximations. The off-shell vertex, which
has a much more complicated structure than the free vertex, 
thus cannot be extracted from experimental data, but should instead
be {\it consistently} calculated within the framework of a microscopic theory.
Such a calculation will yield the dressed off-shell vertices and the 
concommitant contact terms. The proper interpretation of future high precision 
measurements of intermediate energy processes depends crucially on our ability 
to carry out such consistent calculations in realistic microscopic models.

\bigskip  

\leftline{\bf Acknowledgements}
\medskip\noindent
We would like to thank J.H. Koch
for many helpful discussions and a critical reading of the manuscript.
  This work is supported in part by the Foundation for Fundamental
  Research on Matter (FOM) and the National Organisation for 
  Scientific Research (NWO). GP acknowledges support through
  Human Capital and Mobility Fellowship ERBCHBICT941430.

\appendix

\section{}

Here we present some details of our model calculation.
The integral matrix introduced in Eq.~(\ref{tmatrix}),
\be\label{tes}
   \I = \xara(\Ip,0,0,\Ie) \; ,
\ee
describes the meson-nucleon loop Feynman integrals that appear
in the RHS of Figs.\ \ref{tmatr},\ref{kstrong}. Thus,
\be\label{int1}
\I_{i}(w) = -i\int {d^4k\over (2\pi)^4} {1\over (k^2-\mu_i^2)}
{1\over (\thru{p}+\thru{q}-\thru{k}-m)} \ ,
\ee
where $w=\sqrt{(p+q)^2}$. Eq.\ (\ref{int1}) is made meaningful 
by Pauli-Villars regularization of the meson propagators
\be\label{reg}
{i\over k^2-\mu_i^2} \rightarrow {i\over k^2-\mu_i^2} - 
{i\over k^2-M^2} \; .
\ee
For simplicity, we will use the same cutoff mass, $M$, for both
the $\pi$ and $\eta$ propagators.

We may now 
write $\I_i=m\I_i^{0}+(\thru{p}+\thru{q})\I_i^1$, and
defining $\Delta_i=(w^2+m^2-\mu_i^2)^2-4w^2 m^2$, we find
for the imaginary parts
\bea\label{int2}
	\Im \I_i^0 &=& { 1\over 16\pi }
	{ \sqrt{|\Delta_i|}\over w^2}\ttt - 
	\Bigl[ \mu_{i} \!\rightarrow M \!\Bigr]
		\\
	\Im \I_i^1 &=&  { 1\over 16\pi} 
	{(w^2\!+\!m^2\!-\!\mu_i^2)\over 2w^2}
	{\sqrt{|\Delta_i|}\over w^2}\ttt\ 
	-\Bigl[ \mu_{i}\! \rightarrow M \!\Bigr]
		\nnn \ ,
\eea
while the real parts are given by 
\bea\label{int3}
	\Re \I_i^0 &=& {1 \over 16\pi^2 } \int_0^1 dx \ln \vert 
        w^2 x^2 + \beta_i x+ m^2\vert
       - \Bigl[ \mu_i \rightarrow M  \Bigr]\\
	\Re \I_i^1 &=& { 1 \over 16\pi^2}
	\int_0^1 dx x \ln \vert w^2 x^2 + 
           \beta_i x+ m^2\vert -\Bigl[ \mu_i 
	\rightarrow M \Bigr] \nnn \ ,
\eea
with $\beta_i\equiv (\mu_i^2-w^2-m^2)$. We now make the choice 
$\lep=\lpe=\sqrt{\lpp\lee}$ which
considerably simplifies the formulae, rendering $\T$ of the form
\be
\T_{ij}=\L_{ij}{1\over a + (\thru{p}+\thru{q}) b} \ ,
\ee 
with 
\bea\label{pente}
a &\equiv& 1 - m (\lpp \I_{\pi}^0 + \lee \I_{\eta}^0)\nnn \\
b &\equiv& - (\lpp \I_{\pi}^1 + \lee \I_{\eta}^1) \ .
\eea
Using the projection operators defined in Eq.\ (\ref{prop}), 
$T$ may be written as
\be\label{tmat1}
\T={ \L \over a+wb } P_{+} + { \L \over a-wb } P_{-} \; .
\ee
As our T-matrix has no $x=\cos \theta$ dependence, it is easy to
show using the standard partial wave projections that $P_{+}$ projects
only into the $f^{0+}$ partial wave and $P_{-}$ only into the
$f^{1-}$ partial wave. The formalism for meson-nucleon scattering 
is well-known \cite{IZ} and need not be repeated here.
We only mention that phase space factors must be included in the
T-matrix, \equ(tmat1), that is,  in terms of the partial waves, 
$f^{l\pm}_{ij}$, it is the object $T_{ij}=\sqrt{|q_i| |q_j|} f^{l\pm}_{ij}$ 
(no sum over $i,j$) that satisfies the simple unitarity 
equation, (\ref{ena}). Including phase space factors we thus obtain
\bea
	T^{S11}_{ij} &=& {1\over 8\pi w(wb+a)}
			  {\cal F}_{ij}^{+}\label{f1}\\
	T^{P11}_{ij} &=&  {1\over 8\pi w(wb-a)}
			  {\cal F}_{ij}^{-}\label{f2} \ ,
\eea
where ${\cal F}^{\pm} = F^{\pm} \L F^{\pm}$, with $F^{\pm}$ = diag
$\Bigl[\sqrt{|q_{\pi}|(E_{\pi}\pm m)}$,  
$\sqrt{|q_{\eta}|(E_{\eta}\pm m)}\Bigr]$, {\it i.e.}
\be\label{f}
{\cal F}^{\pm} = \xara( f_\pi^\pm,   \sqrt{f_\pi^\pm f_\eta^\pm},
                        \sqrt{f_\pi^\pm f_\eta^\pm}, f_\eta^\pm 
                      ) \ .
\ee
Here $f_\pi^\pm$ = $\lpp |q_{\pi}| (E_{\pi}\pm m)$ (and similarly for
$f_\eta^\pm$) and  the $+(-)$ sign is associated with 
$\T^{S11}(\T^{P11})$, respectively. We now show that unitarity is satisfied.
Notice that ${\cal F^\pm}$ is of the form ${\cal F^\pm}_{ij}=
\sqrt{h_i^\pm h_j^\pm}$
 and thus $({\cal F}^\pm)^2={\rm Tr}({\cal F^\pm}){\cal F^\pm}$.
Using $(w^2+m^2-{\mu}_i^2)=2wE_i$ it is easy to inspect that 
Eqs.~(\ref{int2},\ref{pente}) imply
\bea\label{exi1}
 -8\pi w\; \Im(a+wb) &=& f_\pi^+ + f_\eta^+ \nnn\\
\Rightarrow -8\pi w \;\Im(a+wb){\cal F^+}&=&{\rm Tr}({\cal F^+}){\cal F^+}=
({\cal F}^+)^2\nonumber\\
\Rightarrow -{1\over 8\pi w}{\Im(a+wb)\over |a+wb|^2}{\cal F^+} &=&
{1\over (8\pi w)^2}{1\over |a+wb|^2}({\cal F}^+)^2\nonumber\\
\Rightarrow \Im(T^{S11})&=& T^{S11} (T^{S11})^\dagger\label{exi22} \ ,
\eea
and analogously for $T^{P11}$. Notice that
for $w\ge M+m$ the extra terms  $[ \mu_i 
	\rightarrow M]$
contribute to the RHS of Eq.~(\ref{int2}). As a result, 
Eq.~(\ref{exi1}) is spoiled and unitarity violated.
However, we will take $M$ large enough so that this 
violation of unitarity is of no practical consequence.

Let us next consider the unitarity constraint for the strong 
half-off-shell vertex  $\Gamma^5 u(p)$ given by Eq.~({\ref{strongf}). 
Writing $\T$ in terms of the unitary T-matrix
\be\label{tmat2}
\T = 8\pi w \left[ (F^{+})^{-1}T^{S11}(F^{+})^{-1} P_{+}
	          +(F^{-})^{-1}T^{P11}(F^{-})^{-1} P_{-} \right] \ ,
\ee
and using the definition of $\T$, \equ(tmatrix), we obtain
%%%  Thus, Eq.~(\ref{strongf}) may be written as
\bea
{\Gamma^5 u(p)}  &=& \g5 (1-\L\I)^{-1}\L\L^{-1}G u(p)
\nonumber \\
  &=& 
	\g5 \T \L^{-1} Gu(p)  \\
  &=&   8\pi w \g5 \left[
	(F^{+})^{-1}T^{S11}(F^{+})^{-1} P_{+}
	+ (F^{-})^{-1} T^{P11}(F^{-})^{-1} P_{-} 
        \right]\L^{-1} G u(p)  \nonumber \\
  &=& 
	8\pi w  \left[ (F^{+})^{-1}T^{S11}(F^{+})^{-1} P_{-}
	+ (F^{-})^{-1} T^{P11}(F^{-})^{-1} P_{+} 
	\right] \L^{-1} G\g5 u(p) \nonumber
	\ ,
\eea
where we have used $\g5 P_{\pm}$ = $P_{\mp}\g5$.
Using Eq. (\ref{split}) we find
\bea\label{lala}
	K(+w) &=&
			8\pi w  
			(F^{-}(w))^{-1} T^{P11}(w) 
			(F^{-}(w))^{-1} \L^{-1}G \nonumber \\
	K(-w) &=&
			 8\pi w 
			 (F^{+}(w))^{-1}T^{S11}(w)
			 (F^{+}(w))^{-1} \L^{-1} G 
			 \ .
\eea
Thus, $K(w)$ is related to the P-wave on-shell
amplitude $T^{P11}(w)$ and $K(-w)$ to the S-wave on-shell T-matrix
amplitude  $T^{S11}(w)$. Taking the imaginary part of 
both sides we obtain
\bea
	\Im K(+w) &=&  8\pi w  (F^{-})^{-1} 
	           \Im (T^{P11}) (F^{-})^{-1} \L^{-1} G
				\nonumber \\
	      &=&  8\pi w  (F^{-})^{-1} 
                  (T^{P11})^{\dagger} T^{P11} (F^{-})^{-1}\L^{-1} G
				 \nonumber \\
	      &=&  8\pi w (F^{-})^{-1} 
                   (T^{P11})^\dagger F^{-} (F^{-})^{-1} 
	           T^{P11} 
		   (F^{-})^{-1} \L^{-1} G \nnn \\
               &=&  (F^{-})^{-1} (T^{P11})^\dagger
                   F^{-} K(+w) \label{sss1} \\
               &=& (F^{-})^{-1} (T^{P11}) F^{-}
                    K^*(+w) \nonumber \ ,
\eea
where, in the last step, we have taken the complex conjugate of 
\equ(sss1) and used the fact that $T$ is symmetric so as to 
cast \equ(sss1) in the form of \equ(imktk). Similar equations 
are found  for $K(-w)$ with $F^{-}$ $\rightarrow$ $F^+$ and
$T^{P11}$ $\rightarrow$ $T^{S11}$. 
Notice that we have tacitly assumed that $\L$ has an inverse, 
but we have chosen $\L$ such that this is not the case. However, 
we have explicitly  checked that Eq. (\ref{kun}) remains valid.

\section{}

Here we study the effect of field transformations
on the absorptive (imaginary) part of the strong form factor. 
To do this, we must go beyond tree level. Ideally, we would 
like to perform the transformation, \equ(Dyson), 
to the one-channel version of our 
model, Eq.~(\ref{tria}), checking that the off-shell form factor
shows a different asymptotic behavior while remaining invariant on-shell.
To first order in $\beta$ we obtain
\bea\label{new}
L' = L_f &-& i(g+2m\beta)\bar\psi\g5\phi\psi 
+ \beta \bar\psi\g5(\thru\partial\phi)\psi \nnn\\
&+& {\lambda\over 2}\bar\psi\phi^2 \psi
+i\beta\lambda\bar\psi\g5\phi^3\psi 
%%% + {\cal O}(\beta^2)
 \ ,
\eea
where $L_f$ represents the free (kinetic) part of the Lagrangian. However, 
the equivalence theorem (representation independence of on-shell form 
factors) only holds if {\it all} diagrams to a given order are included.
In particular, diagrams that we have omitted because they do not contribute
to the imaginary part, as for example  diagrams with meson loops that dress
the on-shell nucleon as well as reducible diagrams with closed loops, have
to be included as well. Unfortunately, that means that it is impossible to 
make a non-perturbative comparison since we would have to solve both 
theories exactly, without being able to restrict ourselves to an infinite 
subset of  diagrams as in the previous section. 

We can still, however, make a 
perturbative comparison. First of all, we can check to 
${\cal O}(\beta\lambda)$ that the {\it on-shell} form factors are the same
between $L$ and $L'$. That will
provide an example of the representation {\it in}-dependence of {\it on}-shell 
form factors beyond the tree level result of section II. To show this, we 
need to take into account all diagrams in Fig.\ \ref{tadp}. Notice that
diagram (f) is present only in the transformed Lagrangian, Eq.~(\ref{new}).
The comparison is most easily made by examining how the terms proportional to
the ``tadpole'' integral 
\be
   \It\equiv -i \int {d^4k\over (2\pi)^4} {1\over k^2-\mu^2}
\ee
compare for the on-shell matrix element (hence the $\bar u(p')$
spinor to the left as well) between the two models. From the transformed 
Lagrangian, $L'$, we obtain for the tadpole graphs
\bea
   (d')&=& (d) + {\beta\lambda \It \over 2} 
         \bar u(p')\g5 (2m+\thru q){1\over\thru p-m} u(p)\nnn\\
       &=& (d) - {\beta\lambda \It \over 2} \bar u(p')\g5 u(p) \label{l1d}\\
   (e')&=& (e) + {\beta\lambda \It \over 2} 
	\bar u(p') {1\over\thru{p}'-m} \g5 (2m+\thru q)u(p)\nnn\\
       &=& (e) - {\beta\lambda \It \over 2} \bar u(p')\g5 u(p) \label{l1e}\\
   (f') &=& 3 \beta \lambda \It  \bar u(p')\g5 u(p) \label{l1f} \ .
\eea
%%% We see that the terms with the propagators cancel between $(d+e)$
%%% and $(d'+e'+f')$. 
The $\beta$-dependent contribution from the ($b'$) and ($c'$) graphs
can also be cast in terms of $\It$ by using the Dirac equation for the
on-shell spinors
\bea\label{hjhj}
(b')&=& i \lambda
         \int {d^4k\over (2\pi)^4} {i\over k^2-\mu^2}
     \bar u(p'){i\over \thru{p} - \thru{k} -m} \nnn\\
      && \qquad  \qquad  \times \; \g5(g+2m\beta -\beta \thru{k})
    u(p)\nnn\\
    &=& (b) -i\beta\lambda \int {d^4k\over (2\pi)^4}{1\over k^2-\mu^2}
     \bar u(p'){1\over \thru{p} - \thru{k} -m}\nnn\\ 
&& \qquad  \qquad  \times \; (m-\thru{p} + \thru{k})\g5 u(p) \nnn\\
    &=& (b) - \beta\lambda \It \bar u(p')\g5 u(p) \ , 
\eea
and similarly for $(c')$. From (\ref{l1d}-\ref{hjhj}) it is clear
that the overall coeficient of the $\beta$-dependent terms 
$\lambda \It \bar u(p')\g5 u(p)$ vanishes
\be 
 %%% \beta\lambda \It \bar u(p')\g5 u(p)
 \Bigl( -{1\over 2}- {1\over 2} +3 -1 -1  \Bigr)  = 0 \ .
\ee
That completes the proof of on-shell invariance. What about the imaginary 
part? The on-shell form factor has no imaginary part.
For the half-off-shell ($p^2=m^2$) form factor the tadpole 
contributions are real. The ($b'$) contribution is also real, 
since, for $w=m$, $\Delta$ = $-\mu^2(4m^2-\mu^2)<0$ (cf. \equ(int2)).
Thus, the only diagrams that can generate an imaginary part are 
$(c')$  and $(c)$ (obtained from $(c')$ by taking $\beta\rightarrow 0$).
 We find
\be\label{cprime}
(c)' = \g5\Bigl\{ [g+2 m \beta] \I+\beta \J \Bigr\}\lambda u(p) \ ,
\ee
where $\J$ is a Feynman integral resulting from the ``pseudovector''
coupling and is defined analogously to $\I$ (Eq.~(\ref{int1}) 
and Fig.\ \ref{kstrong})
\bea
\J(p') &=& -i\int {d^4k\over (2\pi)^4} {1\over k^2-\mu^2}\thru{k}
         {1\over \thru{p}+\thru{q}-\thru{k}-m} \nnn\\
       &=&   -i \int {d^4k\over (2\pi)^4}{(\thru p'-m)\over k^2-\mu^2}
             {(\thru{p}'-\thru{k}+m)\over (p'-k)^2-m^2}\nnn\\
       &&  \qquad\qquad +i \int {d^4k\over (2\pi)^4}
           {1\over k^2-\mu^2} \nnn\\
           \Rightarrow \J(p') 
       &=& (\thru{p}'-m)\I(p') -\It \label{int2p}\ ,
\eea
where $\It$ is a real c-number ({\it i.e.}, 
independent of the off-shell variable $w$),
and therefore does not contribute to the once-subracted
dispersion relation.
%%%
%%% that on-shell cancels the closed loop 
%%% contributions as we have  shown above. 
%%%
Thus,
\be\label{cprime2}
(c)' = \biggl\{ \Bigl(g-\beta[\thru p '- m]\Bigr)\I 
      -\beta \It\biggr\}\g5 u(p) \ .
\ee
We clearly see that the same off-shell operator, $(\thru p '- m)$, 
multiplies {\it both} the real and imaginary part of the integral $\I$.
We conclude that the imaginary part of the off-shell form factor
shows a higher-power asymptotic behavior in $w$ that is 
representation-dependent, in agreement with our arguments based
on unitarity.

As a last exercise that clarifies the points made in this work,
consider the half-off-shell vertex function in the following 
two-channel Lagrangians:
\bea
   L_1 = L_{ps} &+& {1\over 2}\bar\psi\F\L\F\psi \label{ggg1} \\
   L_2 = \tilde L &+& {1\over 2}\bar\psi\F\L\F\psi \label{ggg2} \ .
\eea
Here $L_{ps}$ is given by Eq.~(\ref{tria1}) and $\tilde L$
by Eq.~(\ref{triaaa3}). Keeping ${\cal O} (\beta)$ terms only (where
$\beta$ is now a two-component vector like $G$)
$L_1$ is our original Lagrangian, Eq.~(\ref{tria}), whereas the 
second is a different ``model'', {\it not} unitarily-equivalent 
to $L_1$, but nevertheless generating the {\it same} T-matrix
(in the sense of Fig.\ \ref{tmatr}).
To ${\cal O}(G,\beta)$, the off-shell form factor generated by
$L_2$ is (cf. Eq.\ (\ref{strongf}))
\be\label{kab} 
\Gamma_2^5 u(p) = \g5{1\over 1-\L\I} 
 \biggl\{  G +\beta (\thru{p}'+m) -\beta \L\It \biggr\} u(p) \ .
\ee
Projecting as in Eq. (\ref{prop}) we see that, since the 
imaginary part comes solely from the  $1-\L\I$ term, 
(related to the T-matrix) the 
same line of arguments leading to Eq.~(\ref{lala})
shows that {\it both} $K_a,\;a\in\{1,2\}$, satisfy
\be
\Im(K_a)=F^{-1}T F K_a^* \ ,
\ee
with   the {\it same} T-matrix. 
As with $K_1(w)$, we renormalize $K_2(w)$ such that it
is equal to the (physical) $G_{\pi N N}$ coupling at $w=m$, {\it i.e.}
the two form factors $K_1(w)$ and $K_2(w)$ are equal at
the on-shell point $w=m$. However, they have a different
asymptotic behavior in the off-shell variable and therefore require 
a different number of subtractions in the dispersion relation. Thus,
this example shows that knowledge of the T-matrix cannot
uniquely determine the off-shell form factor.

%\end{document}

%%%%%%%%%%%%%%%%%%%%%%%%%%%%%%%%%%%%%%%%%%%%%
%%% \begin{figure}[htb]
%%% \begin{center}
%%% \mbox{\epsfig{file=vertex.ps,width=0.9\textwidth,angle=0}}
%%% \caption[dummy]{ The pion-nucleon (strong) vertex 
%%%         $\Gamma^5(p',p)$, with $p'=p+q$.
%%%         In the most general case all particles may be off-shell.}
%%% \label{vertex}
%%% \end{center}
%%% \end{figure}
%%% \vfill\eject
%%%%%%%%%%%%%%%%%%%%%%%%%%%%%%%%%%%%%%%%%%%%%
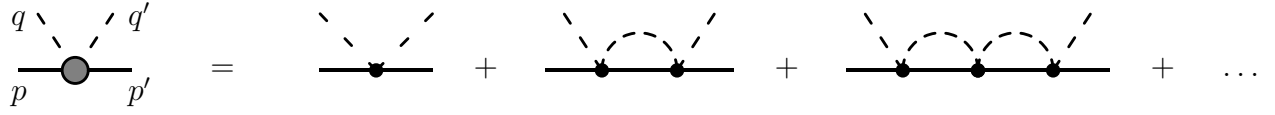
\begin{figure}[htb]
\begin{center}
\noindent
\setlength{\unitlength}{1mm}
\allinethickness{0.4mm}

\newsavebox{\halfcircle}
\savebox{\halfcircle}(0,0)
%\thicklines
{\put(0,0){\arc{10}{3.142}{3.427}}
 \put(0,0){\arc{10}{3.718}{3.998}}
 \put(0,0){\arc{10}{4.284}{4.570}}
 \put(0,0){\arc{10}{4.855}{5.141}}
 \put(0,0){\arc{10}{5.426}{5.712}}
 \put(0,0){\arc{10}{5.994}{6.283}}
% \multiput(-7.5,0)(15,0){2}{\circle*{1}}
 \put(-5.0,0){\circle*{1.5}}
 \put(+5.0,0){\circle*{1.5}}
}

\begin{picture}(175,50)
%\thicklines
\put(0,0){\makebox(20,50)[bl]{
\put(2.5,10){\line(1,0){15}}
\put(10,10){\dashline{2}(0,0)(-5,7.5)}
\put(10,10){\dashline{2}(0,0)( 5,7.5)}
\put(1.5,6){$p$}
\put(17,6){$p'$}
\put(1.5,16){$q$}
\put(17,16){$q'$}
\filltype{shade}
\put(10,10){\circle*{3.5}}
\filltype{black}
}}

\put(28,9.2){$=$}
\put(63,9.2){$+$}
\put(103,9.2){$+$}
\put(153,9.2){$+$}

\put(40,0){\makebox(20,50)[bl]{
\put(2.5,10){\line(1,0){15}}
\put(10,10){\circle*{1.5}}
\put(10,10){\dashline{2}(0,0)(-7.5,7.5)}
\put(10,10){\dashline{2}(0,0)(7.5,7.5)}
}}

\put(70,0){\makebox(30,50)[bl]{
\put(2.5,10){\line(1,0){25}}
\put(15,10){\usebox{\halfcircle}}
\put(10,10){\dashline{2}(0,0)(-5,7.5)}
\put(10,10){\dashline{2}(10,0)(15,7.5)}
}}

\put(110,0){\makebox(40,50)[bl]{
\put(2.5,10){\line(1,0){35}}
\put(15,10){\usebox{\halfcircle}}
\put(25,10){\usebox{\halfcircle}}
\put(10,10){\dashline{2}(0,0)(-5,7.5)}
\put(10,10){\dashline{2}(20,0)(25,7.5)}
}}

\put(160,0){\makebox(10,50)[bl]{
           \put(2.5,9.2){$\dots$}}}

\end{picture}
% \mbox{\epsfig{file=tmatr.ps,width=0.9\textwidth,angle=0}}
\caption[dummy]{ Diagrams contributing to the on-shell meson-nucleon T-matrix
        (Eq.~(\ref{tmatrix}) in text). The dashed lines denote either
        a pion or an eta meson and ($\bullet$) stands for a $\L$-type
        coupling.
}
\label{tmatr}
\end{center}
\end{figure}
%%% \vfill\eject
%%%%%%%%%%%%%%%%%%%%%%%%%%%%%%%%%%%%%%%%%%%%%
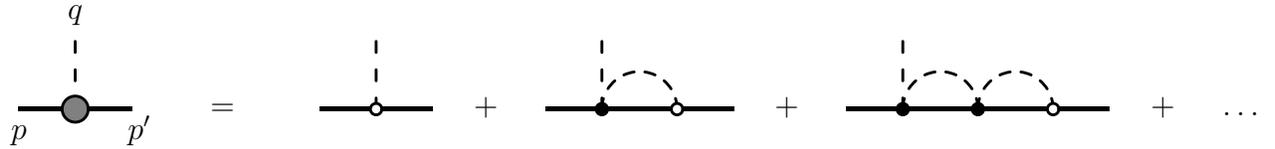
\begin{figure}[htb]
\begin{center}
\setlength{\unitlength}{1mm}
\allinethickness{0.4mm}

%%% \newsavebox{\halfcircle}
%%% \savebox{\halfcircle}(0,0)
%%% \thicklines
%%% {\put(0,0){\arc{10}{3.142}{3.427}}
%%%  \put(0,0){\arc{10}{3.718}{3.998}}
%%%  \put(0,0){\arc{10}{4.284}{4.570}}
%%%  \put(0,0){\arc{10}{4.855}{5.141}}
%%%  \put(0,0){\arc{10}{5.426}{5.712}}
%%%  \put(0,0){\arc{10}{5.994}{6.283}}
% \multiput(-7.5,0)(15,0){2}{\circle*{1}}
%%%  \put(-5.0,0){\circle*{1.5}}
%%%  \put(+5.0,0){\circle*{1.5}}
%%% }

\newsavebox{\samehalfcircle}
\savebox{\samehalfcircle}(0,0)
%\thicklines
{\put(0,0){\arc{10}{3.142}{3.427}}
 \put(0,0){\arc{10}{3.718}{3.998}}
 \put(0,0){\arc{10}{4.284}{4.570}}
 \put(0,0){\arc{10}{4.855}{5.141}}
 \put(0,0){\arc{10}{5.426}{5.712}}
 \put(0,0){\arc{10}{5.994}{6.283}}
 \put(-5.0,0){\circle*{1.5}}
%% \put(+5.0,0){\circle{1.5}}
}

\newsavebox{\mydash}
\savebox{\mydash}(0,0)
{\put(0,7.5){\dashline{2}(0,0)(0,9)}
%%  \put(0,0){\path(0,0)(1,0)}}
%%  \put(2,0){\path(0,0)(1,0)}} 
%%  \put(4,0){\path(0,0)(1,0)}}
%%  \put(6,0){\path(0,0)(1,0)}}
%%  \put(8,0){\path(0,0)(1,0)}}
}

\begin{picture}(175,50)
%\thicklines
\put(0,0){\makebox(20,50)[bl]{
\put(2.5,10){\line(1,0){15}}
\put(10,10){\usebox{\mydash}}
\put(1.5,6){$p$}
\put(17,6){$p'$}
\put(9,22){$q$}
\filltype{shade}
\put(10,10){\circle*{3.5}}
\filltype{black}
}}

\put(28,9.2){$=$}
\put(63,9.2){$+$}
\put(103,9.2){$+$}
\put(153,9.2){$+$}

\put(40,0){\makebox(20,50)[bl]{
\put(2.5,10){\line(1,0){15}}
\put(10,10){\usebox{\mydash}}
\filltype{white}
\put(10,10){\circle*{1.5}}
\filltype{black}
}}

\put(70,0){\makebox(30,50)[bl]{
\put(2.5,10){\line(1,0){25}}
\put(15,10){\usebox{\samehalfcircle}}
\put(10,10){\usebox{\mydash}}
\filltype{white}
\put(20,10){\circle*{1.5}}
\filltype{black}
}}

\put(110,0){\makebox(40,50)[bl]{
\put(2.5,10){\line(1,0){35}}
\put(15,10){\usebox{\samehalfcircle}}
\put(25,10){\usebox{\samehalfcircle}}
\put(10,10){\usebox{\mydash}}
\filltype{white}
\put(30,10){\circle*{1.5}}
\filltype{black}
}}

\put(160,0){\makebox(10,50)[bl]{
           \put(2.5,9.2){$\dots$}}}

\end{picture}
%\mbox{\epsfig{file=kstrong.ps,width=0.9\textwidth,angle=0}}
\caption[dummy]{ Diagrams contributing to the half-off-shell ($p^2=m^2$,
        $p'^2\ne m^2$)  meson-nucleon (strong) form factor
        (Eq.~(\ref{strongf}) in text). The dashed lines may denote 
        either a pion or an eta meson, ($\bullet$) stands for a $\L$-type 
        coupling and (o) for a $G$-type coupling.
}
\label{kstrong}
\end{center}
\end{figure}
\vfill\eject
%%%%%%%%%%%%%%%%%%%%%%%%%%%%%%%%%%%%%%%%%%%%%
\begin{figure}[htb]
\begin{center}
\mbox{\epsfig{file=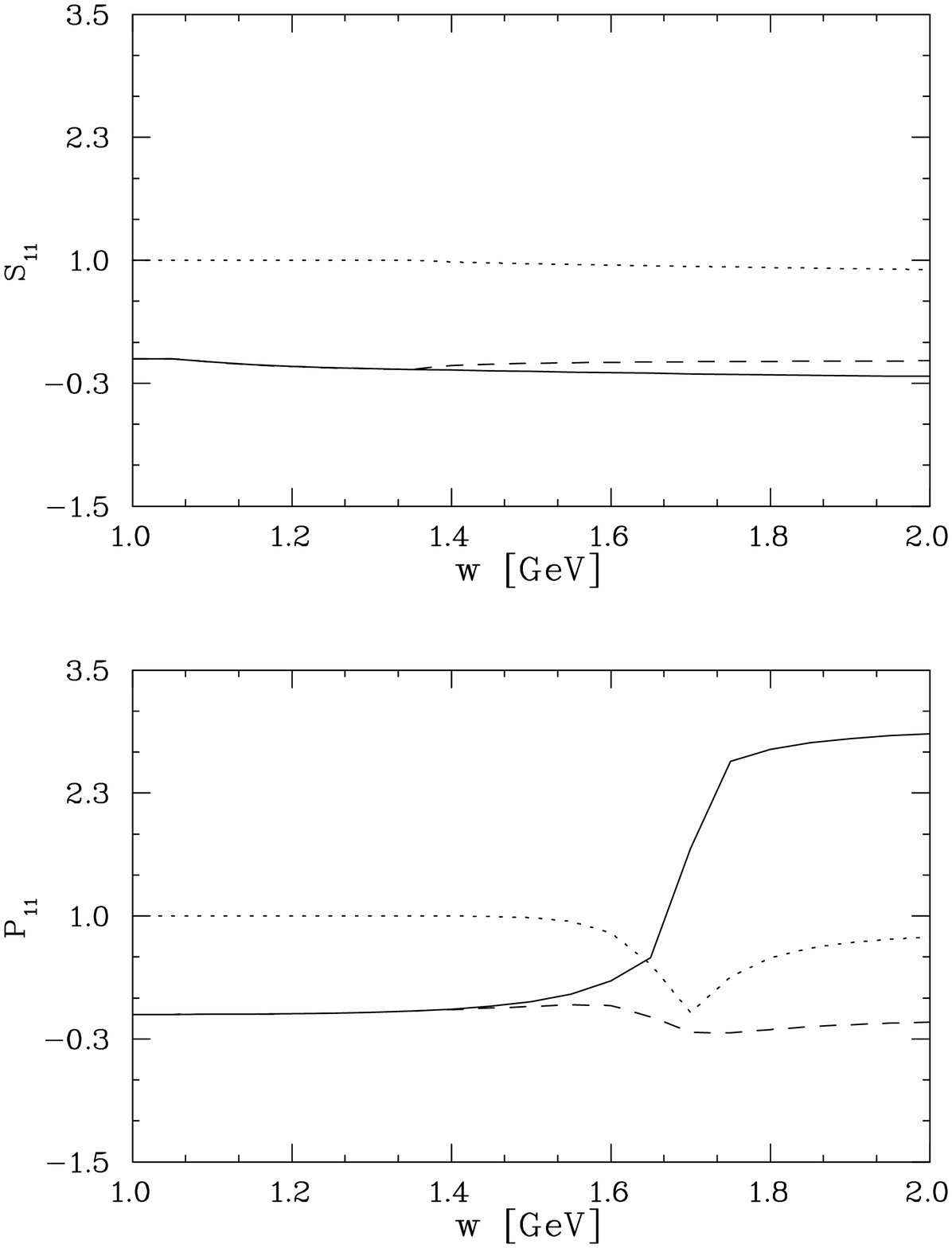,width=0.9\textwidth,angle=0}}
\caption[dummy]{Inelasticities (dotted lines), the pion-nucleon phase shift 
in radians (solid lines) and the phase of the pion-nucleon form 
        factor (dashed lines) for the $S_{11}$ and
        $P_{11}$ channels in our model with parameters $\lpp=0.5$ 
        MeV$^{-1}$,
        $\lee=0.8$ MeV$^{-1}$, $g_{\pi}=-g_{\eta}=2$. 
}
\label{phsh}
\end{center}
\end{figure}
\vfill\eject
%%%%%%%%%%%%%%%%%%%%%%%%%%%%%%%%%%%%%%%%%%%%%
%%% \begin{figure}[htb]
%%% \begin{center}
%%% \mbox{\epsfig{file=conv.ps,width=0.9\textwidth,angle=0}}
%%% \caption[dummy]{The convergence of the subtracted dispersion relation integral
%%%         {\it i.e.}, the RHS of Eq.~(\ref{sub}) 
%%%         for the pion-nucleon form factor ($\bullet$) and the 
%%%         eta-nucleon
%%%         form factor (o). The LHS of the same equation is shown as well. 
%%%         The cutoff has been chosen as $M=10$ GeV.}
%%% \label{conv}
%%% \end{center}
%%% \end{figure}
%%% \vfill\eject
%%%%%%%%%%%%%%%%%%%%%%%%%%%%%%%%%%%%%%%%%%%%
\begin{figure}[htb]
\begin{center}
\mbox{\epsfig{file=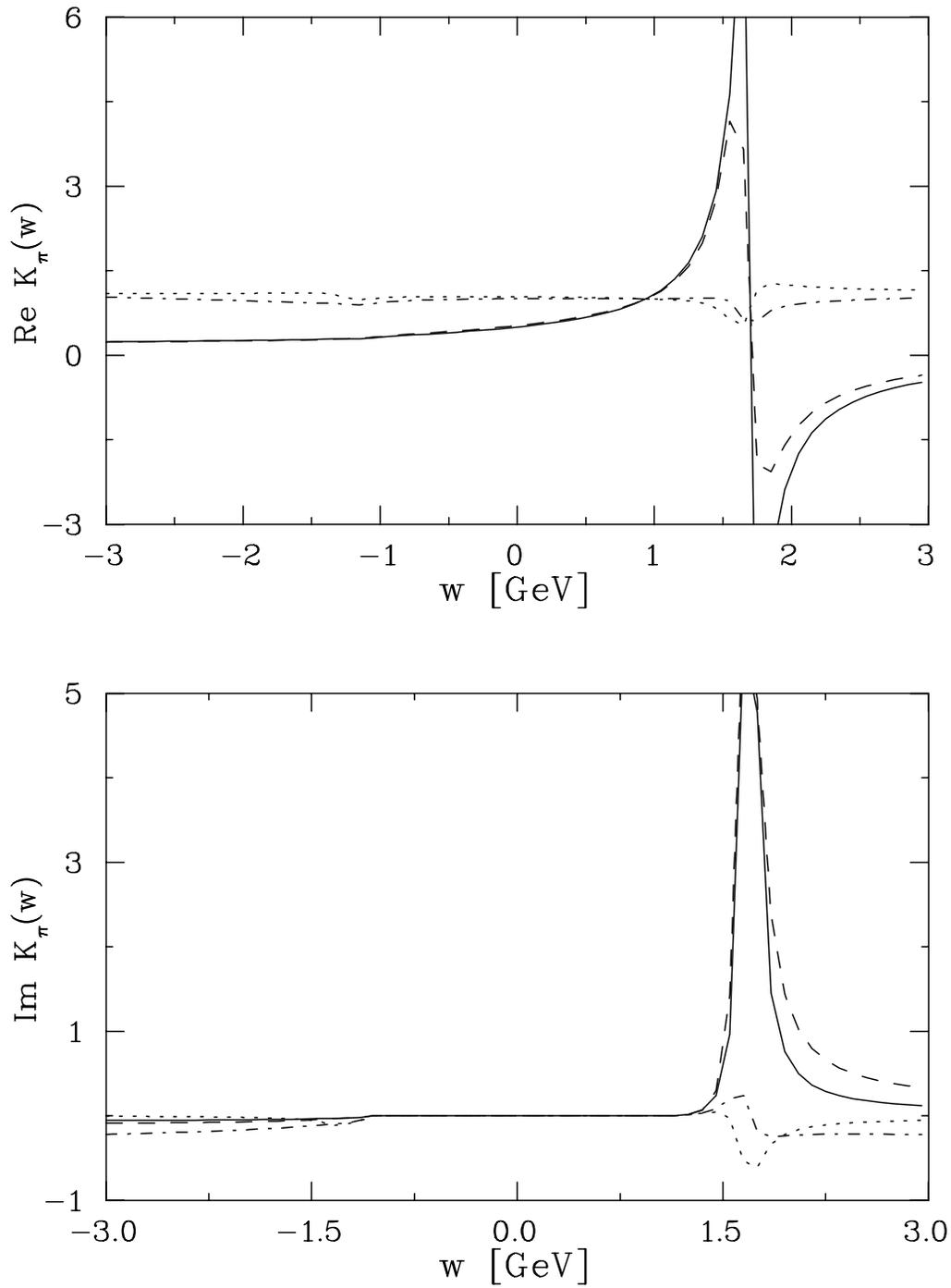,width=0.9\textwidth,angle=0}}
\caption[dummy]{Various approximations for determining the pion form 
        factor phase (solid = threshold, dot-dashed = Goldberger-Treiman),
         compared with the model prediction for the phase for two choices of
         $G$-couplings corresponding to   
       $K_\eta(m)/K_\pi(m)= -1$ (dots) and
       $K_\eta(m)/K_\pi(m)= +1$ (dashes).
       
}
\label{allap}
\end{center}
\end{figure}
\vfill\eject
%%%%%%%%%%%%%%%%%%%%%%%%%%%%%%%%%%%%%%%%%%%%
\begin{figure}[htb]
\begin{center}
\mbox{\epsfig{file=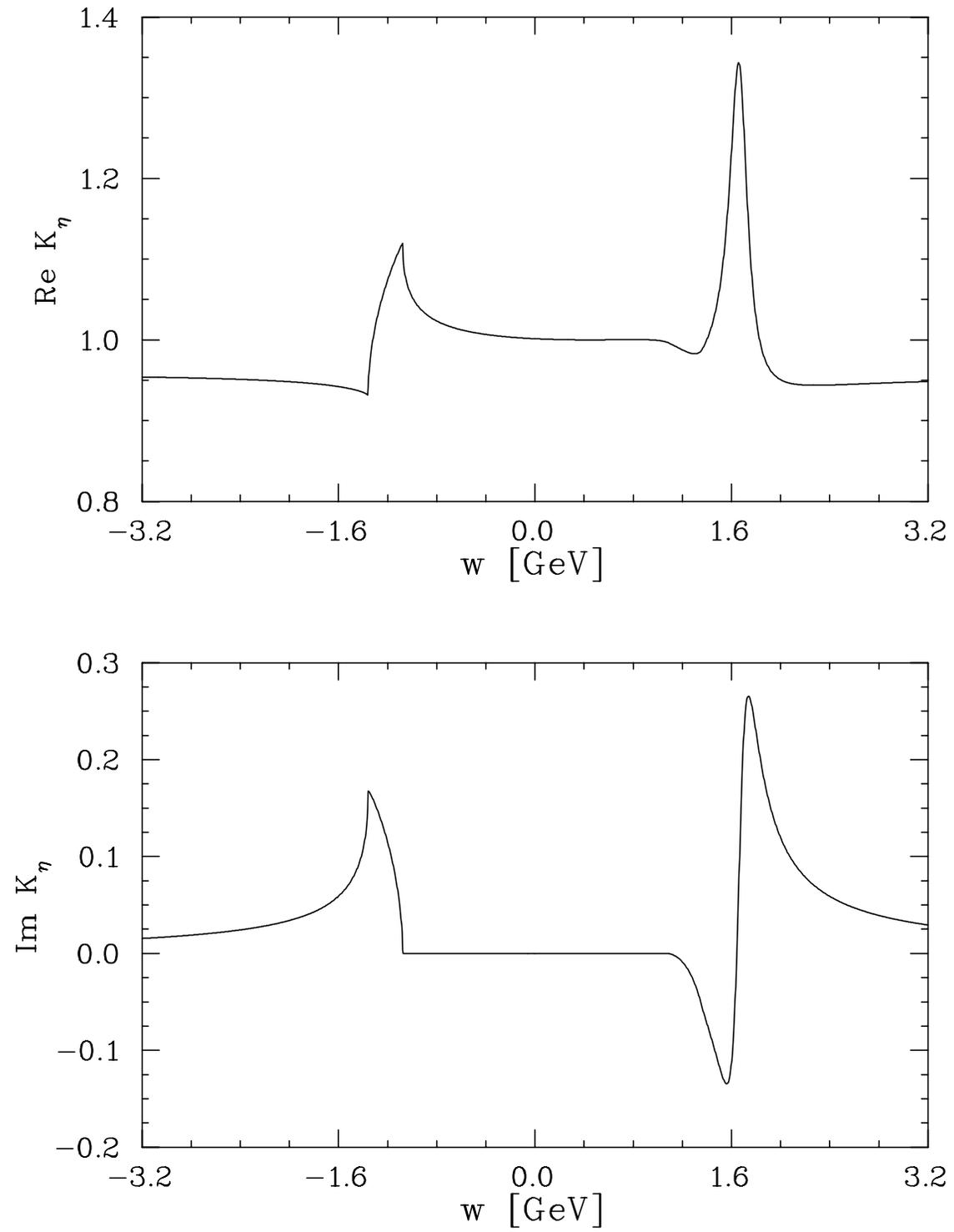,width=0.9\textwidth,angle=0}}
\caption[dummy]{The eta-nucleon form factor.
 Parameters  as in  Fig. \ \ref{phsh}
}
\label{keta}
\end{center}
\end{figure}
\vfill\eject
%%%%%%%%%%%%%%%%%%%%%%%%%%%%%%%%%%%%%%%%%%%%
\begin{figure}[htb]
\begin{center}
\setlength{\unitlength}{1mm}
\allinethickness{0.4mm}

\newsavebox{\halfcircleII}
\savebox{\halfcircleII}(0,0)
%\thicklines
{\put(0,0){\arc{10}{3.142}{3.427}}
 \put(0,0){\arc{10}{3.718}{3.998}}
 \put(0,0){\arc{10}{4.284}{4.570}}
 \put(0,0){\arc{10}{4.855}{5.141}}
 \put(0,0){\arc{10}{5.426}{5.712}}
 \put(0,0){\arc{10}{5.994}{6.283}}
% \multiput(-7.5,0)(15,0){2}{\circle*{1}}
% \put(-5.0,0){\circle*{1.5}}
% \put(+5.0,0){\circle*{1.5}}
}

\newsavebox{\tadpole}
\savebox{\tadpole}(0,0)
{
%%%%%%%%%%%%%%%%%%%%%%%%%%%%%%%%%%%%%%%%%%%%%%%%%%%%
%%% \put(0,0){\arc{10}{0 }{.5}}
%%% \put(0,0){\arc{10}{1.0 }{1.5}}
%%% \put(0,0){\arc{10}{2.0 }{2.5}}
%%% \put(0,0){\arc{10}{3.142}{3.427}}
%%% \put(0,0){\arc{10}{3.718}{3.998}}
%%% \put(0,0){\arc{10}{4.284}{4.570}}
%%% \put(0,0){\arc{10}{4.855}{5.141}}
%%% \put(0,0){\arc{10}{5.426}{5.712}}
%%% \put(0,0){\arc{10}{5.994}{6.283}}
%%% if wanna change tadpole shape execute tadpole.f
%%%\input junk.tex
\put(0,0){\arc{10}{  0.}{   0.241661}}
\put(0,0){\arc{10}{   0.483322}{   0.724983}}
\put(0,0){\arc{10}{   0.966644}{    1.20830}}
\put(0,0){\arc{10}{    1.44997}{    1.69163}}
\put(0,0){\arc{10}{    1.93329}{    2.17495}}
\put(0,0){\arc{10}{    2.41661}{    2.65827}}
\put(0,0){\arc{10}{    2.89993}{    3.14159}}
\put(0,0){\arc{10}{    3.38325}{    3.62491}}
\put(0,0){\arc{10}{    3.86658}{    4.10824}}
\put(0,0){\arc{10}{    4.34990}{    4.59156}}
\put(0,0){\arc{10}{    4.83322}{    5.07488}}
\put(0,0){\arc{10}{    5.31654}{    5.55820}}
\put(0,0){\arc{10}{    5.79986}{    6.04152}}
\put(0,0){\arc{10}{    6.28319}{    6.52485}}
%%%%%%%%%%%%%%%%%%%%%%%%%%%%%%%%%%%%%%%%%%%%%%%%%%%%

}
\newsavebox{\newdash}
\savebox{\newdash}(0,0)
{  \put(0,10){\dashline{2}(0,0)(0,10)}
%%  \put(0,0){\path(0,0)(1,0)}}
%%  \put(2,0){\path(0,0)(1,0)}} 
%%  \put(4,0){\path(0,0)(1,0)}}
%%  \put(6,0){\path(0,0)(1,0)}}
%%  \put(8,0){\path(0,0)(1,0)}}
}

\begin{picture}(175,100)
\put(20,40){\makebox(40,40)[bl]{
\put(10,20){\line(1,0){20}}
%%% \put(20,20){\usebox{\newdash}}
\put(20,20){\dashline{2}(0,0)(0,12)}
\put(19,10){$a$}
}}

%%% \put(28,9.2){$=$}

\put(70,40){\makebox(40,40)[bl]{
\put(7.5,20){\line(1,0){25}}
\put(20,20){\usebox{\halfcircleII}}
\put(25,20){\dashline{2}(0,0)(0,12)}
\put(19,10){$b$}
}}

\put(120,40){\makebox(40,40)[bl]{
\put(7.5,20){\line(1,0){25}}
\put(20,20){\usebox{\halfcircleII}}
\put(15,20){\dashline{2}(0,0)(0,12)}
\put(19,10){$c$}
}}

\put(20,0){\makebox(40,40)[bl]{
\put(7.5,20){\line(1,0){25}}
\put(15,25){\usebox{\tadpole}}
\put(25,20){\dashline{2}(0,0)(0,12)}
\put(19,0){$d$}
}}

\put(70,0){\makebox(40,40)[bl]{
\put(7.5,20){\line(1,0){25}}
\put(25,25){\usebox{\tadpole}}
\put(15,20){\dashline{2}(0,0)(0,12)}
\put(19,0){$e$}
}}

\put(120,0){\makebox(40,40)[bl]{
\put(7.5,20){\line(1,0){25}}
\put(20,15){\usebox{\tadpole}}
\put(20,20){\dashline{2}(0,0)(0,12)}
\put(19,0){$f$}
}}

\end{picture}
\bigskip
% \mbox{\epsfig{file=tadp.ps,width=0.9\textwidth,angle=0}}
\caption[dummy]{Feynman diagrams for Lagrangian $L'$ of Eq.~(\ref{new})
       contributing the form factor at order $\beta\L$.
}
\label{tadp}
\end{center}
\end{figure}
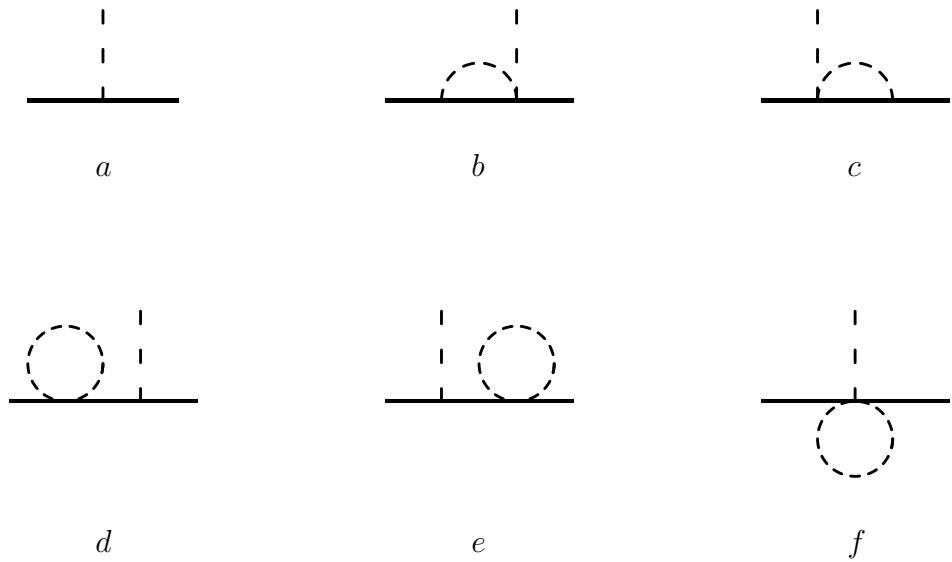
%%%%%%%%%%%%%%%%%%%%%%%%%%%%%%%%%%%%%%%%%%%%


\begin{thebibliography}{99}

\bibitem{Bincer} A. Bincer, Phys. Rev. {\bf 118}, 855 (1960).

\bibitem{hare1} M.G. Hare and G. Papini, Phys. Rev. D {\bf 4},
684 (1971).
M.G. Hare, Ann. Phys. {\bf 74}, 595 (1972).

\bibitem{Bos} J.W. Bos, Ph.D. thesis, University of Amsterdam (1993).

\bibitem{bluv} R.E. Bluvstein, A.A. Cheshkov, and V.M. Dubovik,
Nucl. Phys. {\bf B64}, 407 (1973). R.E. Bluvstien and V.M. Dubovik,
Phys. Lett. {\bf B45}, 345 (1973).

\bibitem{suura} H. Suura and L.M. Simmons, Phys. Rev. {\bf 148},
1579 (1966).

\bibitem{Epstein} G.N. Epstein, Phys. Lett. {\bf B79}, 195 (1978).

\bibitem{Treiman} M. Goldberger and S. Treiman, Phys. Rev. {\bf 111},
354 (1958).

\bibitem{LSZ} H. Lehmann, K. Symanzik and W. Zimmerman,
               Nuovo Cim. {\bf 1}, 1 (1955).

\bibitem{Coleman} S. Coleman, J. Wess, and B. Zumino, Phys. Rev. 
{\bf 177}, 2239 (1969).

\bibitem{Kamefuchi} S. Kamefuchi, L. O'Raifearteigh and A.~Salam,
          Nucl. Phys. {\bf 28}, 529 (1961).

\bibitem{Coleman_book}
            Sidney Coleman, {\it Aspects of Symmetry},
            (Cambridge University Press, 1985) Chapter 2, 
            pages 39, 40;
            for a formal proof see J.H. Borchers, Nuov. Cim.
             {\bf 15}, 784 
           (1960, in German).

\bibitem{Rudy} T.E. Rudy, H.W. Fearing, and S. Scherer,
            Phys. Rev. C {\bf 50}, 447 (1994).

\bibitem{Stefan_Compton} S. Scherer and H.W. Fearing, Phys. 
 Rev. C {\bf 51}, 359 (1995).

\bibitem{hoe} G. H\"{o}hler in Landoldt-B\"{o}rnstein (H. Schooper,
ed., Springer, Berlin, 1983) Vol. 9 b2.

\bibitem{gas} J. Gasser, H. Leutwyler, M.P. Locher, and M.E. Sainio,
Phys. Lett. {\bf B213}, 85 (1988).

\bibitem{Kazes} E. Kazes, Nuov. Cim. {\bf 13}, 1226 (1959).

\bibitem{deForest} T. de Forest, Jr. Nucl. Phys. {\bf A392}, 
 239 (1983).

\bibitem{gross}F. Gross and D.O. Riska, Phys. Rev. C {\bf 36},
1928 (1987).

\bibitem{koch} S. Scherer and J. Koch, Nucl. Phys. {\bf A534},
461 (1991).

\bibitem{nyman} E.~Nyman, Nucl. Phys. {\bf A154}, 97 (1970).

\bibitem{fish} W.E. Fischer and P. Minkowski, Nucl. Phys.
		{\bf B36}, 519 (1972).

\bibitem{Naus} H.W.L. Naus and J.H. Koch, Phys. Rev. C {\bf 36},
2459 (1987).

\bibitem{TT} P.C. Tiemeijer and J.A. Tjon, 
                          Phys. Rev. C {\bf 42}, 599 (1990).

\bibitem{mac} X. Song, J.P. Chen and J.S. McCarthy, Z. Phys. A {\bf 341},
	275 (1992).

\bibitem{bos2} J.W. Bos and J.H. Koch, Nucl Phys. {\bf A563}, 539 (1993).

\bibitem{Friar} J.L. Friar and B.F. Gibson, 
   Phys. Rev. C {\bf 15}, 1779 (1977).

\bibitem{Harolds_suggestion} We thank Harold Fearing for 
              suggesting this form of presentation to us.

\bibitem{BJ} J.D. Bjorken and S.D. Drell, \begin{it} Relativistic
Quantum Fields,\end{it} (McGraw-Hill, New York 1965), p. 281.

\bibitem{others} Our separable interaction approach is similar to
                 that of D.J Ernst {\it et al.}, 
               Phys. Rev. C {\bf 10}, 1708 
                 (1974). For a  more realistic T-matrix calculation 
                   in a unitary, coupled, three-channel model 
                see M. Batinic {\it et al.}, 
              Phys. Rev. C {\bf 51} (1995).

\bibitem{IZ} C. Itzykson and J. Zuber, 
\begin{it} Quantum Field Theory,\end{it}
(McGraw-Hill, New York, 1985).

\bibitem{tiator} M. Benmerrouche and N.C. Mukhopadhyay, Phys. Rev.
Lett. {\bf 67}, 1070 (1991). 
                 L. Tiator, C. Benhold and S.S. Kamalov,
                 Nucl. Phys. {\bf A580}, 455 (1994).

\end{thebibliography}
\end{document}